 \documentclass[12pt,superscriptaddress,preprint,nofootinbib]{revtex4-1}
\usepackage{amssymb,graphics,graphicx,tabularx,color,epstopdf}
\usepackage{psfrag}
\usepackage{physics} % includes amsmath, plus many physics typesetting commands
\usepackage{amssymb}
\usepackage{amsmath}
\usepackage{bm} %bold math symbols
\usepackage{graphicx}
\usepackage{wrapfig}
\usepackage{natbib}
\usepackage[colorlinks=true,urlcolor=blue,bookmarks=true,citecolor=blue,breaklinks=true,linkcolor=black]{hyperref}
\usepackage{epstopdf}
\usepackage{enumerate}
\usepackage{xcolor}
\usepackage{float}
\usepackage{longtable}
\usepackage{gensymb}
\usepackage{braket}

\begin{document}
\title{Spontaneous Breaking of Gauge Groups to Discrete Symmetries}
\author{Bradley L. Rachlin\footnote{bradleyrachlin@gmail.com}$^{1}$ and Thomas W. Kephart\footnote{tom.kephart@gmail.com}}
\affiliation{  Department of Physics and Astronomy, Vanderbilt University, Nashville, Tennessee 37235, USA}
\date{\today}

\begin{abstract}
Many models of beyond Standard Model physics connect
flavor symmetry with a discrete group. Having
this symmetry arise spontaneously  from a gauge theory maintains compatibility with
quantum gravity and can be used to systematically prevent anomalies. We minimize a number of Higgs  potentials
that break gauge groups to discrete symmetries of interest, and examine their scalar mass spectra.\\

%\vspace{5cm}

%\begin{center}
%{\color[rgb]{1,0.5,0} {\Huge DRAFT}}
%\end{center}
\end{abstract}
\maketitle

\section{ Introduction}

The standard model (SM) does not explain quark and lepton masses, nor does it explain how quarks and leptons mix. The most studied and perhaps the most aesthetic approach to parameterizing the masses and mixing angles is to extend the SM by a discrete group $\Gamma$ into whose irreducible representations (irreps) the standard model particles are assigned. Many choices of this discrete flavor symmetry have been tried.  As expected, larger groups can typically provide a fuller description of flavor physics, but there are examples of relatively small nonabelian discrete groups like $A_4$ and $T'$ that are somewhat more economical. Here we take an agnostic approach as to the choice of discrete group and study a representative set of examples that have been used in model building.\\

Notable early extensions of the standard model with discrete symmetries include the work of Pakvasa and Sugawara \cite{Pakvasa:1977in} who used $\Gamma=S_3$ and focused on the quark sector, as well as Ma and collaborators \cite{Ma:2001dn,Babu:2002dz} who used $\Gamma=A_4$ to describe the lepton sector. Many other choices for $\Gamma$ have been used in model building, several of which will be discussed below. For an early brief review of possible discrete groups that can be used for SM extensions see \cite{Frampton:1994rk}. Recent extensive reviews with more complete and up to date bibliographies are also available. See for instance \cite{Altarelli:2010gt,Ishimori:2010au,King:2013eh,King:2014nza}.

Extending the SM by a discrete group is not without its perils. Global discrete symmetries are violated by gravity \cite{Krauss:1988zc}, the discrete group can be anomalous \cite{Luhn:2008sa}, unwanted cosmic defects can be produced \cite{Hindmarsh:1994re}, etc. To avoid as many of these problems as possible the most expedient approach is to gauge the discrete symmetry, i. e., extend the SM by a continuous group $G$ in such a way that no chiral anomalies are produced. Then one breaks this gauge group to the desired discrete group, $G \rightarrow \Gamma$, where now $\Gamma$ is effectively anomaly free and avoids problems with gravity.

Various examples of Lie groups breaking to discrete groups have been discussed in the literature, but only in a few cases have the details of the minimization of the scalar potential and the extraction of the scalar spectrum been investigated. Here we plan to include these important details for many of the discrete groups of interest via the
following procedure:\\
(i) First we provide irreps of $G$ that contain trivial $\Gamma$ singlets. 
These results are summarized in the Appendix. \\
(ii) Next we set up scalar potentials $V$ with scalars in one of these irreps. \\
(iii) Then we find a vacuum expectation value (VEV) via the Reynolds operator \cite{Reynolds} (similar to the perhaps more familiar Molien series \cite{Molien}) that can break $G$ to $\Gamma$. \\
(iv) Next we minimize $V$ to show that the VEV indeed does properly break the symmetry. \\
(v) Finally, we provide the spectrum of scalar masses at the $\Gamma$ level after the breaking. 
Our calculations are carried out with Mathematica and checked by hand where practical.

Many of the methods we employ were developed in work by  Luhn \cite{Luhn:2011ip} and by Merle and Zwicky \cite{Merle:2011vy},   where some of the results summarized here can be found. 
We believe our results will be of interest to many model builders, since it will allow them to include the minimal set of scalars necessary to break a gauge symmetry to a discrete symmetry of interest. A few examples that go beyond the minimal set of scalars are also included, where the symmetry breaking is carried out from a nonminimal $G$ irrep or a non-minimal $G$.

\section{ Lie Group Invariant Potentials} 
Our task in this section is to construct Higgs potentials invariant under  Lie groups $G$ for specific irreps. But first we must see which irreps are suitable for spontaneous symmetry breaking (SSB), i.e., irreps whose decompositions include a trivial singlet of the desired subgroup
$ \Gamma \subset G$ to which we hope to break. Using the  Mathematica package {\it decomposeLGreps} \cite{Fallbacher:2015pga} along with GAP to generate the groups \cite{GAP4}, one can easily produce tables of branching rules from Lie group irreps to subgroup irreps   and find such singlets. We have done this for a number of cases and have included them in a short appendix  for convenience and to make the paper self contained.

\subsection{ Gauge group irreps containing discrete gauge singlets} The discrete groups $\Gamma$ we will discuss and the gauge groups where they can be minimally embedded   are $A_4, S_4, A_5 \subset SO(3)$; $Q_6,T', O', I'\subset SU(2)$; and $T_7,\Delta(27), PSL(2,7)\subset SU(3)$. 

These discrete groups can also be embedded non-minimally. For example, we include the case $A_4\subset SU(3)$. Minimal and non-minimal embedding of other discrete groups can be handled in a way similar to what is discussed here, and we hope that the examples we discuss are sufficiently informative to aid in other cases. 

To spontaneously break $G$ to $\Gamma$ with some irrep $R$ of $G$, it is necessary that $R$ contains a trivial $\Gamma$ singlet. It is straightforward to look at the decomposition of $R$ from $G$ to $\Gamma$ to make this determination. The decomposition can be carried out by standard techniques starting from character tables. Since it is the character tables that are usually provided in the literature, we here provide an appendix with the tables of decompositions of the first few irreps of $SO(3)$, $SU(2)$, and $SU(3)$ to discrete groups of interest. For example, as one can see in Table \ref{tab:4} of the Appendix, the 7 and 9 dimensional irreps of $SO(3)$ have trivial $A_4$ singlets, therefore these irreps are candidates for the scalar potential that allows the spontaneous symmetry breaking $SO(3)\rightarrow A_4$. 

\subsection{ $SO(3)$ Potentials}

We will begin our study of SSB by starting with relatively simple examples and then proceed to more sophisticated cases. But first, a note on cubic terms in the potential; a general renormalizable potential has quadratic, cubic, and quartic terms, but the cubic terms tend to significantly complicate the analysis. We exclude these terms for simplicity by imposing a $Z_2$ symmetry (or, like in some cases, they vanish upon summation), so the following potentials are actually $SO(3)\times Z_2$ invariant. (The $Z_2$ symmetry can be avoided by including the cubic terms or by gauging it too.) The effect of including the cubic terms is studied for some cases where the analysis is tractable in Section \ref{scalesspectra}. We now proceed to our first example, the breaking pattern $SO(3) \rightarrow A_4$.
\subsubsection{$A_4$}
We begin by constructing an $SO(3)$ invariant potential
\footnote{Group Theory Comments:
The tetrahedral group $A_{4} \subset SO(3)$ has double-valued representations that 
correspond to single-valued representations of the binary (double) tetrahedral group $T' \subset SU(2)$. 
As $SO(3)$ is not a subgroup of $SU(2)$, likewise $A_{4}$ is not a subgroup of $T'$~\cite{Frampton:2009pr}. 
Hence, besides the  irreps of $T'$ that are coincident 
with those of $A_{4}$, it has three additional spinor doublet-like irreps.
The relationships between $S_4$ and $O'$ and between  $A_{5}$ and $I'$  are similar.}.
As stated above, which irrep 
\newpage
\noindent
we use depends on the discrete subgroup of interest. For example, if we want to break to the tetrahedral group $A_4$, which has been used to describe  the tri-bimaximal neutrino mixing pattern \cite{Altarelli:2010gt} \cite{Berger:2009tt} and co-bimaximal  mixing \cite{Ma:2015fpa}, we look at Table \ref{tab:4} and see that the lowest dimensional irrep we can use is the \textbf{7}. For references to other recent work with $A_4$ models see \cite{Altarelli:2008bg,King:2011ab,Ferreira:2013oga}. In terms of the fundamental \textbf{3} of $SO(3)$, we   obtain a \textbf{7} as a direct product of three \textbf{3}s.

\begin{equation}
\mathbf{3} \times \mathbf{3} \times \mathbf{3} = \mathbf{1}+ 3\cdot\mathbf{3} +2\cdot\mathbf{5}+\mathbf{7}
\end{equation}  
 
\noindent
This product gives a generic rank 3 tensor with 27 independent components. To isolate 
the \textbf{7}, we take only the totally symmetric part, which reduces the number of components from 27 to 10, giving the symmetric tensor $S_{ijk}$. Then, using the fact that the Kroenecker delta $\delta_{ij}$, is an invariant of the fundamental irrep of $SO$ groups (for a discussion of Lie group invariant tensors see \cite{Cvitanovic:1976am}) we subtract off the three traces, $\sum_j^3 \delta_{jk}S_{ijk}$, i=1,2,3 , to obtain the traceless symmetric tensor $T_{ijk}$, which is our 7 dimensional $SO(3)$ irrep. As mentioned above, the most general renormalizable potential is constructed from the independent quadratic, cubic, and quartic contractions of this tensor. In this case there are two quartic terms,  but notice that all the cubic terms, which necessarily include the anti-symmetric Levi-Civita Tensor, $\epsilon_{ijk}$ vanish upon summation. Hence the potential for the \textbf{7} is
\begin{equation}\label{eq:V7}
V_7 = -m^2 \,T_{ijk}T_{ijk}+\lambda \,(T_{ijk}T_{ijk})^2 +\kappa \,\,T_{ijm}T_{ijn}T_{kln}T_{klm}
\end{equation}

In subsequent sections we find a vector (in a particular basis) pointing in the $A_4$
 direction, then minimize the potential and find the mass eigenstates and show that they can all be positive which implies the minimum is stable. Minimization implies certain constraints on the coupling constants must be satisfied as will be discussed. We proceed in analogous fashion for other $G \rightarrow \Gamma$ cases, but first we will collect all the potentials we need for the purpose.
 
 \subsubsection{$S_4$}
To break to the octahedral group, $S_4$, we see from Table \ref{tab:10} that the lowest irrep we can use is the \textbf{9}. From examining Kroenecker products, we see that we must begin with the direct product of four \textbf{3}s. Similar to the results in the previous subsection, we take the symmetric part of this rank 4 tensor, $S_{ijkl}$, which reduces the number of components to 15. We then subtract off the six trace elements, $\,\sum \delta_{kl}S_{ijkl}$, to obtain the desired 9-component tensor. The associated potential is
\begin{equation}\label{eq:V9}
\begin{split}
V_9 = -&m^2 \,T_{ijkl}T_{ijkl}+\lambda \,(T_{ijkl}T_{ijkl})^2 +\kappa \,T_{ijkl}T_{ijkp}T_{mnop}T_{mnol}\\
+&\rho\,T_{ijkl}T_{ijop}T_{mnop}T_{mnkl}+\tau \, T_{ijkl}T_{ijmn}T_{kmop}T_{lnop}
\end{split}
\end{equation}
For examples where the octahedral group has been used to build models see \cite{Hagedorn:2006ug,Hagedorn:2010th}.
 
 \subsubsection{$A_5$} 
Another subgroup of interest, which has been used in a number of recent models  \cite{Everett:2008et,Feruglio:2011qq,Ding:2011cm,Chen:2010ty},  is $A_5$. From Table \ref{tab:3} we see that the \textbf{13} is the lowest irrep that contains a trivial $A_5$ singlet. Again starting from the fundamental $SO(3)$ triplet one can show that the Kroenecker product of six \textbf{3}s is needed to get an irrep of this dimension. The symmetric part of this rank 6 tensor, $S_{ijklmn}$ has 28 independent components, which is then reduced to 13 by subtracting off the 15 trace elements, $\,\sum \delta_{mn}S_{ijklmn}$. The potential is constructed in a  fashion similar to the $A_4$ case.
\begin{equation}\label{SO3_13}
\begin{split}
V_{13} =-&m^2\,T_{ijklmn}T_{ijklmn}+\lambda \,(T_{ijklmn}T_{ijklmn})^2 + \kappa \,T_{ijklmn}T_{ijklmt}T_{opqrsn}T_{opqrst}\\
+&\rho \,T_{ijklmn}T_{ijklst}T_{opqrmn}T_{opqrst} +\tau \,T_{ijklmn}T_{ijkrst}T_{opqlmn}T_{opqrst} 
\end{split}
\end{equation}

\subsection{ $SU(2)$ Potentials}
We now proceed in a similar vein  to construct $SU(2)$ invariant potentials. In fact, for the odd dimensional (real) representations, invariants must be constructed from   triplets which furnish an unfaithful representation of $SU(2)$. As such the true symmetry of the theory is not given by the potential alone and must be determined from the specifics of the model, i.e., from the full Lagrangian. In the following cases, the omission of the cubic terms means the potentials have a $SU(2)\times U(1)$ symmetry, where the $U(1)$ is a phase. This phase can also be gauged and then broken if necessary to avoid problems with global symmetries, or in some cases cubic terms can be added that do not respect the $U(1)$.

\subsubsection{$Q_6$}

If we want to break to $Q_6$ we see from Table \ref{tab:1} that the lowest dimensional irrep we can use is the \textbf{5}. However, as explained in \cite{Frampton:1994xm,Frampton:1999hk}, this irrep will actually break to the continuous subgroup $Pin(2)$. So we must look at the next lowest irrep with a trivial $SU(2)$ singlet, the \textbf{7}. We cannot break with a real \textbf{7} as in Eq.\eqref{eq:V7} because there are no triplet representations of $Q_6$ that can be used to find a VEV in the unfaithful SO(3) representation. Thus we must use the complex \textbf{7}, which has the same potential as needed for the $T'$ case which is given in Eq.\eqref{complex7} below.

\subsubsection{$T'$}
To break  from $SU(2)$ to $T'$, the binary tetrahedral group, we see from Table \ref{tab:2} that the smallest $SU(2)$ irrep we can use is the \textbf{7}. Since we must construct it from   triplets the potential is   the same as in equation \eqref{eq:V7}. The VEVs will also be the same.

Another possibility is to do  the breaking to $T'$ with a complex \textbf{7}, which can be thought of as a pair of real \textbf{7}s. We can now build our representation out of the fundamental doublets of $SU(2)$, where we  get  the \textbf{7} by taking the direct  product of six \textbf{2}s and isolating the tensor symmetric on all indices. The potential is
\begin{equation}\label{complex7}
\begin{split}
V_{7_c}=-&m^2 \,T_{ijklmn}T^{ijklmn}+\lambda \,(T_{ijklmn}T^{ijklmn})^2+\kappa \,T_{ijklmn}T^{ijklmt}T_{opqrst}T^{opqrsn}\\
+&\rho \,T_{ijklmn}T^{ijklst}T_{opqrst}T^{opqrmn}+\tau \,T_{ijklmn}T^{ijkrst}T_{opqrst}T^{opqlmn}
\end{split}
\end{equation}
where the indices now run from 1 to 2. All cubic terms have vanished upon summation.
$T'$ models are economical and have been used to explain both quark and lepton sector parameters  \cite{Frampton:1994rk, Aranda:1999kc,Chen:2007afa,Frampton:2007et,Frampton:2007et,Frampton:2008bz,Frampton:2010uw,Natale:2016xob,Carone:2016xsi}. A more complete set of recent $T'$ model references can be found in 
\cite{Carone:2016xsi}.

\subsubsection{$O'$}
To break to from $SU(2)$ to $O'$, the binary octahedral group, we see from Table \ref{tab:11} that the smallest $SU(2)$ irrep we can use is the \textbf{9}. As in the $S_4$ example, we can construct our potential from   triplets so the potential is  the same as in equation \eqref{eq:V9} and the VEVs will again be the same.

We can also consider the case of a complex \textbf{9} and build the representation out of $SU(2)$ doublets. We obtain the \textbf{9} through the symmetric product of eight \textbf{2}s. The potential is

\begin{equation}\label{complex9}
\begin{split}
V_{9_c}=-&m^2 \,T_{ijklmnop}T^{ijklmnop}+\lambda \,(T_{ijklmnop}T^{ijklmnop})^2+\kappa \,T_{ijklmnop}T^{ijklmnox}T_{qrstuvwx}T^{qrstuvwp}\\
+&\rho \,T_{ijklmnop}T^{ijklmnwx}T_{qrstuvwx}T^{qrstuvop}+\tau \,T_{ijklmnop}T^{ijklmvwx}T_{qrstuvwx}T^{qrstunop}\\
+&\sigma \,T_{ijklmnop}T^{ijkluvwx}T_{qrstuvwx}T^{qrstmnop}
\end{split}
\end{equation}
$O'$ is maximal in $SU(2)$, so the proper SSB is assured for a VEV that is $O'$ invariant.

\subsubsection{$I'$}

The final $SU(2)$ breaking case we consider is $I'$, the binary icosahedral group, which has been used in both three and four family extensions of the SM \cite{Everett:2010rd,Chen:2011dn}. Here the lowest $SU(2)$ irrep we can use is the real \textbf{13}, which yields the same potential as we used for $A_5$ (Eq. \eqref{SO3_13}).

Alternatively for the case of a complex \textbf{13} we see that it is given by the symmetric product of twelve \textbf{2}s. The potential has seven quartic invariants, and the first few terms are

\begin{equation}\label{V13*}
\begin{split}
V_{13_c} =&-m^2\,T_{abcdefghijkl}T^{abcdefghijkl}+\lambda \,(T_{abcdefghijkl}T^{abcdefghijkl})^2 \\
+&\kappa \,T_{abcdefghijkl}T^{abcdefghijkx}T_{mnopqrstuvwx}T^{mnopqrstuvwl} + ...\\
\end{split}
\end{equation}
Potentials for higher tensors can be cumbersome to write, so let us introduce a new notation to deal with them. For instance for the potential for the \textbf{13}, let us define
$$T_{12a}\cdot T^{12a}= T_{abcdefghijkl}T_{abcdefghijkl},$$
and
$$(T_{11a}\cdot T^{11a})^b_c(T_{11a}\cdot T^{11a})^c_b= T_{abcdefghijkl}T^{abcdefghijkx}T_{mnopqrstuvwx}T^{mnopqrstuvwl},  $$ etc. Specifically we write $na$ for the collection of indices $a_1a_2a_3...a_n$, etc.
Then the full potential is for the complex \textbf{13} takes the form
\begin{equation}\label{V13*}
\begin{split}
V_{13_c} =&-m^2\,T_{12a}\cdot T^{12a}+\lambda \, (T_{12a}\cdot T^{12a})^2
+\kappa (T_{11a}\cdot T^{11a})^b_c(T_{11a}\cdot T^{11a})^c_b \\
+&\rho      \,(T_{10a}\cdot T^{10a})^{2b}_{2c}(T_{10a}\cdot T^{10a})^{2c}_{2b}
 + \tau \,(T_{9a}\cdot T^{9a})^{3b}_{3c}(T_{9a}\cdot T^{9a})^{3c}_{3b} 
 +\nu   \,(T_{8a}\cdot T^{8a})^{4b}_{4c}(T_{8a}\cdot T^{8a})^{4c}_{4b}\\
 +&\sigma  \,(T_{7a}\cdot T^{7a})^{5b}_{5c}(T_{7a}\cdot T^{7a})^{5c}_{5b}
 +\chi \,(T_{6a}\cdot T^{6a})^{6b}_{6c}(T_{6a}\cdot T^{6a})^{6c}_{6b} \\
\end{split}
\end{equation}
This notation is consistent when the tensor $T$ is totally symmetric on all of its indices \footnote{We could write an even more compact notation in generalized dyadic form, e.g., the $\nu$ term would be
$\nu (T:^8T):^4(T:^8T)$ which again defines how the tensor contractions are to be made, but we find this form unnecessary here, but it could be useful for expressions involving more complicated group invariants. Cvitanovic's ``Bird Track'' notation\cite{Cvitanovic:1976am} can also be useful for this purpose}.

Again, since $I'$ is maximal in $SU(2)$, the proper SSB is assured for an $I'$ invariant VEV.
\subsection{ $SU(3)$ potentials}
Similar to the previous section, the omission of cubic terms means that the following potentials have an $SU(3)\times U(1)$ symmetry, where the $U(1)$ can be dealt with as described above.

\subsubsection{$A_4$ and $T_7$}
In addition to SO(3), $A_4$ can originate from a broken $SU(3)$ symmetry. Looking at Table \ref{tab:6} we see that the lowest dimensional irrep containing a trivial $A_4$ singlet is the \textbf{6}, but as explained in \cite{Luhn:2011ip}, neither the \textbf{6}, \textbf{10}, nor $\bf{15'}$ will break $SU(3)$ uniquely to $A_4$, i.e., giving these irreps an $A_4$ VEV will necessarily leave a group larger than $A_4$ unbroken. This leaves us with the \textbf{15} as the smallest irrep that will uniquely break to an $A_4$  subgroup, and the same logic applies to $T_7$. (A variety of $T_7$ models have been proposed, see  \cite{Luhn:2007sy,Kile:2014kya,Vien:2015koa,Vien:2016qbb}.) To obtain a useful form of the $\bf{15}$ we first take the product $\mathbf{3} \times \mathbf{3} \times \mathbf{{\bar 3}}$   in $SU(3)$; then by specifying the part that is symmetric on 2 indices, $S_{ij}^k$, we reduce the number of independent components from 27 to 18. Finally, subtracting off the three traces: $\sum_j^3 \delta_{jk}S_{ij}^{k}, \, i=1,2,3\,$, gives us the desired 15 component tensor. The associated potential \cite{Luhn:2011ip} is

\begin{equation}\label{V15}
\begin{split}
V_{15}=&-m^2 \,T_{ij}^kT_k^{ij}+\lambda \,(T_{ij}^kT^{ij}_k)^2+\kappa \,T_{jm}^iT^{jn}_iT_{ln}^kT^{lm}_k\\
&+\rho \,T_{jm}^iT^{jn}_iT_{kl}^mT^{kl}_n+\tau \,T_{ij}^mT^{ij}_nT_{kl}^nT^{kl}_m+\nu \,T_{jm}^iT^j_{in}T_l^{km}T^{ln}_k
\end{split}
\end{equation}
 
\subsubsection{$\Delta(27)$}
From Table \ref{tab:7} we see that we can use the \textbf{10} to spontaneously break from $SU(3)$ to $\Delta(27)$. We can get to this irrep by taking the product of three triplets and specifying the fully symmetric part of the resulting tensor, which reduces to the desired ten independent components. The potential is 
\begin{equation}\label{V10}
V_{10} = -m^2 \,T_{ijk}T^{ijk}+\lambda \,(T_{ijk}T^{ijk})^2 +\kappa \,\,T_{ijm}T^{ijn}T_{kln}T^{klm}
\end{equation}
where the cubic terms have vanished upon summation. This result can also be found in \cite{Luhn:2011ip}. Examples where $\Delta(27)$ has been used are \cite{Ferreira:2012ri,Vien:2016tmh}.

\subsubsection{$PSL(2,7)$}
Another group that has garnered considerable interest as a flavor symmetry is $PSL(2,7)$ \cite{Chen:2014wiw}. Looking at Table \ref{tab:8} we see that the lowest dimensional irrep of $SU(3)$ we can use to break to $PSL(2,7)$ is the \textbf{15$\bm{'}$}, (Dynkin label [4 0]). To get to a \textbf{15$\bm{'}$} we take the product of four fundamental triplets
\begin{equation}
\mathbf{3} \times \mathbf{3} \times \mathbf{3} \times \mathbf{3} =  3\cdot\mathbf{3} +2\cdot\mathbf{\bar{6}}+3\cdot\mathbf{15}+\mathbf{15\bm{'}}
\end{equation}  
The generic rank 4 tensor has 81 independent components, requiring it be symmetric on all four indices reduces it to $\bf{15'}$ as required. The associated potential is
\begin{equation}\label{V15'}
\begin{split}
V_{15'}=-&m^2 \,T_{ijkl}T^{ijkl}+\lambda \, (T_{ijkl}T^{ijkl})^2\\
+&\kappa \,T_{ijkl}T^{ijkm}T_{mnop}T^{lnop} +\rho \,T_{ijkl}T^{ijmn}T_{mnop}T^{klop}
\end{split}
\end{equation}

Also of interest is the next lowest irrep suitable for breaking from $SU(3)$ to $PSL(2,7)$, the \textbf{28}. We build this irrep by taking the symmetric product of six triplets, giving a fully symmetric rank 6 tensor with 28 components. The associated potential is
\begin{equation}\label{V28}
\begin{split}
V_{28}=-&m^2 \, T_{ijklmn}T^{ijklmn} \, +\lambda \, (T_{ijklmn}T^{ijklmn})^2\\
+&\kappa\,T_{ijklmn}T^{ijklmt}T_{opqrst}T^{opqrsn}+\rho \,T_{ijklmn}T^{ijklst}T_{opqrst}T^{opqrmn}\\
+&\tau \, T_{ijklmn}T^{ijkrst}T_{opqrst}T^{opqlmn}
\end{split}
\end{equation}

\section{ Vaccuum Alignments for Spontaneous Symmetry Breaking}
\subsection{ Vacuua for $SO(3)$ Potentials} 
The invariant tensors from the previous section can be written in terms of a $d$-dimensional orthonormal bases, where $d$ is the number of independent tensor components. To illustrate this consider  the \textbf{5} of $SO(3)$ which is a second rank symmetric traceless tensor $T_{ij}$. It has a basis
\begin{equation}
\begin{split}
&\ket{1}= \frac{1}{\sqrt{2}}(\ket{11}-\ket{22})\\
&\ket{2}= \frac{1}{\sqrt{6}}(\ket{11}+\ket{22}-2\cdot\ket{33})\\
&\ket{3}= \frac{1}{\sqrt{2}}(\ket{12}+\ket{21})\\
&\ket{4}= \frac{1}{\sqrt{2}}(\ket{13}+\ket{31})\\
&\ket{5}= \frac{1}{\sqrt{2}}(\ket{23}+\ket{32})\\
\end{split}
\end{equation}

Where $\ket{ij}$ is the $ij^{th}$ component of the tensor. Using this basis the  matrix form of $T_{ij}$ is
\begin{equation}
T_{ij}=\begin{pmatrix} 
\frac{1}{\sqrt{2}}\ket{1}+\frac{1}{\sqrt{6}}\ket{2}& \frac{1}{\sqrt{2}}\ket{3} & \frac{1}{\sqrt{2}}\ket{4} \\
\frac{1}{\sqrt{2}}\ket{3} & -\frac{1}{\sqrt{2}}\ket{1}+\frac{1}{\sqrt{6}}\ket{2} &  \frac{1}{\sqrt{2}}\ket{5} \\
 \frac{1}{\sqrt{2}}\ket{4} & \frac{1}{\sqrt{2}}\ket{5} & -\sqrt{\frac{2}{3}}\ket{2}
\end{pmatrix}
\end{equation}
With an explicit basis, it now makes sense to look for a d-component vacuum alignment that minimizes the potential and is invariant under the desired discrete subgroup. How do we find this specified direction? First, note that we can express our basis above in polynomial form, assigning component 1 to $x$, 2 to $y$, and 3 to $z$:
\begin{equation*}
\begin{split}
&\ket{1}=\frac{1}{\sqrt{2}}(x^2-y^2)\\
&\ket{2}=\frac{1}{\sqrt{6}}(x^2+y^2-2z^2)\\
&\ket{3}=\frac{1}{\sqrt{2}}(xy+yx)=\sqrt{2}xy\\
&\ket{4}=\sqrt{2}xz\\
&\ket{5}\sqrt{2}yz\\
\end{split}
\end{equation*}
So if we find a polynomial that is invariant under the desired subgroup we can  convert it into a vacuum alignment by expressing it as a vector in terms of these basis functions\cite{Merle:2011vy}. To find a polynomial, $\mathcal{I}(x,y,z)$, invariant under a group $H$, one employs the Reynolds Operator \cite{Reynolds}
\begin{equation} \label{eq:16}
\mathcal{I}(x,y,z)=\frac{1}{|\mathcal{R}(H)|}\sum_{h\in\mathcal{R}(H)}f(h\circ\begin{pmatrix}x\\y\\z\end{pmatrix})
\end{equation}
Where $\mathcal{R}(H)$ is a representation of the group, $|\mathcal{R}(H)|$ is the number of elements in the group, and $f(h\circ\begin{pmatrix}x\\y\\z\end{pmatrix})$ signifies the result of a group element $h$ acting on the vector $(x,y,z)$  and then input into a trial function $f(x,y,z)$. Trial polynomials of the form $x^ny^mz^{d-n-m}$ will typically be most useful in finding invariants of degree $d$. Note we have specified polynomials in three variables here, but we can use the same procedure to find invariants in terms of any number of variables, real or complex.  E.g., in two real dimensions we can find an invariant $\mathcal{I}(x,y)$ with a trial function $f(x,y)$.

\subsubsection{$A_4$}\label{sec:so3a4}
 
As an initial practical example lets examine the symmetry breaking pattern $SO(3) \rightarrow A_4$. The irrep of interest is a $\bf{7}$ which is the symmetric, traceless part of ${\bf 3\times3\times3}$. Expressed it in terms of 7 independent components we have
\begin{equation}
\begin{split}
&\ket{1}= \frac{1}{2}(\ket{111}-\ket{122}-\ket{212}-\ket{221})\\
&\ket{2}= \frac{1}{\sqrt{60}}(3\cdot\ket{111}+\ket{122}+\ket{212}+\ket{221}-4\cdot\ket{133}-4\cdot\ket{313}-4\cdot\ket{331})\\
&\ket{3}= \frac{1}{2}(\ket{222}-\ket{112}-\ket{121}-\ket{211})\\
&\ket{4}= \frac{1}{\sqrt{60}}(3\cdot\ket{222}+\ket{112}+\ket{121}+\ket{211}-4\cdot\ket{233}-4\cdot\ket{323}-4\cdot\ket{332})\\
&\ket{5}= \frac{1}{2}(\ket{333}-\ket{113}-\ket{131}-\ket{311})\\
&\ket{6}= \frac{1}{\sqrt{60}}(3\cdot\ket{333}+\ket{113}+\ket{131}+\ket{311}-4\cdot\ket{223}-4\cdot\ket{232}-4\cdot\ket{322})\\
&\ket{7}= \frac{1}{\sqrt{6}}(\ket{123}+\ket{132}+\ket{213}+\ket{231}+\ket{312}+\ket{321})\\
\end{split}
\end{equation}

Using $xyz$ as a trial polynomial in equation (\ref{eq:16}), ($d=3,\, n=m=1$) gives us back $xyz$ as our invariant polynomial. Expressed in terms of this basis our $A_4$ invariant vacuum alignment is remarkably simple:
\begin{equation} \label{eq:so3a4vev}
v= [0,0,0,0,0,0,1]
\end{equation}

The VEV for spontaneous breaking will be this unit vector multiplied by a constant which minimizes the potential. We must show that this VEV is unique to $A_4$. The gauge group will spontaneously break  to the largest subgroup which leaves that VEV invariant. So $G$ will only break to a desired subgroup, $H$, if there is no other group,  $H'$, which is invariant under the specified VEV and satisfies $H \subset H'\subset G$.
It is difficult to systematically determine which subgroup will be left invariant for a given breaking, and in particular if there is a higher invariance than the desired discrete group, so each case must be considered individually. For the present case we start with the fact that the only groups that contain $A_4$ and are subgroups of $SO(3)$ are $S_4$ and $A_5$. Examining the branching rules for both these groups, one sees that a \textbf{7} of $SO(3)$ does not break to a trivial singlet of either $S_4$ or $A_5$, and thus the largest group left invariant by this VEV must be $A_4$. Hence we have obtained the desired result for the case at hand.

\subsubsection{$S_4$}

For the \textbf{9} of SO(3), it is  more convenient to express our basis in terms of spherical harmonics:

\begin{equation}\label{S4basis}
\begin{split}
&\ket{1}= Y_4^0;
\,\,\, \ket{2}= \frac{i}{\sqrt{2}}(Y_4^1 + Y_4^{-1});
\,\,\, \ket{3}= \frac{1}{\sqrt{2}}(Y_4^1 - Y_4^{-1});
\,\,\, \ket{4}= \frac{1}{\sqrt{2}}(Y_4^2 + Y_4^{-2});\\
&\ket{5}= \frac{i}{\sqrt{2}}(Y_4^2 - Y_4^{-2});
\,\,\, \ket{6}= \frac{i}{\sqrt{2}}(Y_4^3 + Y_4^{-3});
\,\,\, \ket{7}= \frac{1}{\sqrt{2}}(Y_4^3 - Y_4^{-3});\\
&\ket{8}= \frac{1}{\sqrt{2}}(Y_4^4 + Y_4^{-4});
\,\,\, \ket{9}= \frac{i}{\sqrt{2}}(Y_4^4 - Y_4^{-4}).\\
\end{split}
\end{equation}

We find that the polynomial, $x^4+y^4+z^4$ is $S_4$ invariant. Expressed in terms of our basis this is
\begin{equation}\label{vevS4} 
v= [\sqrt{\frac{7}{5}},0,0,0,0,0,0,1,0]
\end{equation}

$S_4$ is also a maximal subgroup of SO(3), so we can be certain our alignment breaks SO(3) uniquely to $S_4$.

\subsubsection{$A_5$}

As mentioned previously, to break from $SO(3)$ to $A_5$ the irrep of interest is the totally symmetric traceless  tensor with 13 independent components contained in ${\bf 3\times3\times3\times3\times3\times3}$. In this case it is again easier (and yields equivalent results) to express the components in terms of spherical harmonics\footnote{One can also use this method for the $A_4$ case, see \cite{Koca:2003jy}.} of degree l\,=\,6,\: $Y_6^m \: (\text{where} \: m=-6,-5...0...5, 6)$. In order to get real basis vectors, we define them as
\begin{equation}
\begin{split}
&\ket{1}= Y_6^0;
\,\,\, \ket{2}= \frac{i}{\sqrt{2}}(Y_6^1 + Y_6^{-1});
\,\,\, \ket{3}= \frac{1}{\sqrt{2}}(Y_6^1 - Y_6^{-1});
\,\,\, \ket{4}= \frac{1}{\sqrt{2}}(Y_6^2 + Y_6^{-2});\\
&\ket{5}= \frac{i}{\sqrt{2}}(Y_6^2 - Y_6^{-2});
\,\,\, \ket{6}= \frac{i}{\sqrt{2}}(Y_6^3 + Y_6^{-3});
\,\,\, \ket{7}= \frac{1}{\sqrt{2}}(Y_6^3 - Y_6^{-3});\\
&\ket{8}= \frac{1}{\sqrt{2}}(Y_6^4 + Y_6^{-4});
\,\,\, \ket{9}= \frac{i}{\sqrt{2}}(Y_6^4 - Y_6^{-4});
\,\,\, \ket{10}= \frac{i}{\sqrt{2}}(Y_6^5 + Y_6^{-5});\\
&\ket{11}= \frac{1}{\sqrt{2}}(Y_6^5 - Y_6^{-5})\;
\,\,\, \ket{12}= \frac{1}{\sqrt{2}}(Y_6^6 + Y_6^{-6});
\,\,\, \ket{13}= \frac{i}{\sqrt{2}}(Y_6^6 - Y_6^{-6}).\\
\end{split}
\end{equation}

We find that a degree six invariant polynomial is $(\frac{(1 + \sqrt{5})^2}{4}  \, x^2 - y^2) (\frac{(1 + \sqrt{5})^2}{4}   \, y^2 - 
   z^2) ( \frac{(1 + \sqrt{5})^2}{4}  \, z^2-x^2) $ \cite{Merle:2011vy}. The associated VEV is proportional to
\begin{equation}\label{vevA5} 
%to be verified
v= [1,0,0,-\sqrt{\frac{21}{2}},0,0,0,-\sqrt{7},0,0,0,\sqrt{\frac{105}{22}},0]
\end{equation}

Because $A_5$ is a maximal subgroup of SO(3), i.e., there is no group $H'$ that nontrivially satisfies $A_5 \subset H'\subset SO(3)$ for any VEV of the $\bf{13}$, and again we can be sure the VEV in eq. (\ref{vevA5}) breaks $SO(3)$ uniquely to $A_5$.

\subsection{ Vacuua for SU(2) Potentials}

\subsubsection{$Q_6$}
For the breaking $SU(2) \rightarrow Q_6$ we use the same basis as with $T'$ above. We find the polynomial $\frac{1}{2}(x^6+y^6)$ is left invariant by $Q_6$, and this leads to a VEV proportional to
\begin{equation}\label{Q6vev}
v=[1,1,0,0,0,0,0,0,0,0,0,0,0,0]
\end{equation}

To make sure we have broken to $Q_6$ and not any larger subgroups, we first note that the \textbf{7} does not break to any $Q_n$ with $n>6$ (see page 6 of \cite{Frampton:1999hk}). The only other larger $SU(2)$ subgroup that can be spontaneously broken with a \textbf{7} is $T'$, but we find that $T'$ has only   one degree six invariant which is given in the subsection above. Therefore, the VEV in Eq. (\ref{Q6vev}) is the  result we were seeking.

\subsubsection{$T'$}
Because $SU(2)$ breaks to  $T'$  from the same real seven dimensional irrep that breaks $SO(3)$ to $A_4$, the potentials are the same and the basis will be the same as in the $A_4$ section above. In addition, the Reynolds operator yields the same polynomial invariant $xyz$, so the VEV is identical.  On the other hand the complex \textbf{7} has a different basis\footnote{Because this is a complex irrep there are actually 14 basis states; the basis states listed are the 7 real   parts of the tensor components, while bases 8 through 14 are the imaginary parts. These conjugate components have been suppressed here since they will always be set to zero at vacuum in order to have a real VEV. This will be the case for most of the complex irreps we consider.}, specifically that of the symmetric tensor with 6 indices.
\begin{equation}
\begin{split}
\ket{1}=&\ket{111111}\\
\ket{2}=&\frac{1}{\sqrt{6}}(\ket{111112}+\ket{111121}+\ket{111211}+\ket{112111}+\ket{121111}+\ket{211111})\\
\ket{3}=&\frac{1}{\sqrt{15}}(\ket{111122} + \ket{111212} + \ket{111221} + \ket{112112} + \ket{112121} + \ket{112211}+\\
&\ket{121112} + \ket{121121} + \ket{121211} + \ket{122111} + \ket{211112} + \ket{211121} + \ket{211211} + \ket{212111} + \ket{221111})\\
\ket{4}=&\frac{1}{\sqrt{20}}(\ket{111222} + \ket{112122} + \ket{112212} + \ket{112221} + \ket{121122} + \ket{121212} + \ket{121221} + \ket{122112} +\\
&\ket{122121} + \ket{122211} + \ket{211122} + \ket{211212} + \ket{211221} + \ket{212112} + \ket{212121} + \ket{212211} +\\
&\ket{221112} + \ket{221121} + \ket{221211} + \ket{222111})\\
\ket{5}=&\frac{1}{\sqrt{15}}(\ket{112222} + \ket{121222} + \ket{122122} + \ket{122212} + \ket{122221} + \ket{211222} + \ket{212122} + \ket{212212} +\\
&\ket{212221} + \ket{221122} + \ket{221212} + \ket{221221} + \ket{222112} + \ket{222121} + \ket{222211}\\
\ket{6}=&\frac{1}{\sqrt{6}}(\ket {122222} + \ket{212222} + \ket{221222} + \ket{222122} + \ket{222212} + \ket{222221}\\
\ket{7}=&\ket{222222}\\
\end{split}
\end{equation}

We find that the polynomial $\frac{1}{2}(xy^5-yx^5)$  is left invariant for this representation and the associated VEV is proportional to
\begin{equation}\label{vevTprime}
v=[0,-1,0,0,0,1,0,0,0,0,0,0,0,0]
\end{equation}

To make sure we have broken to T$'$ we must show that this VEV does not break $SU(2)$ to any larger group. The only $SU(2)$ subgroups that contain T$'$ as a subgroup are I$'$, the binary icosahedral group, and $O'$, the binary octahedral group. Looking at tables of branching rules we see that the \textbf{7} of $SU(2)$ does not contain a trivial singlet of either of these groups, so we can be sure the breaking is to $T'$ as desired.

\subsubsection{$O'$}
Like the other double cover groups, the basis and vacuum direction  for the breaking of $O'$ with a real \textbf{9} of $SU(2)$ will be the same as its $SO(3) \rightarrow S_4$ counterpart  above.

The complex \textbf{9} arises from the basis of the symmetric tensor with 8 doublet indices:
\begin{equation}
\begin{split}
&\ket{1}=\frac{}{}\ket{11111111};
\,\,\,  \ket{2}=\frac{}{}\ket{22222222};
\,\,\, \ket{3}=\frac{1}{\sqrt{8}}(\ket{11111112} + perms);\\
&\ket{4}=\frac{1}{\sqrt{8}}(\ket{22222221} + perms);
\,\,\, \ket{5}=\frac{1}{\sqrt{28}}(\ket{11111122} + perms);\\
&\ket{6}=\frac{1}{\sqrt{28}}(\ket{22222211} + perms);;
\,\,\, \ket{7}=\frac{1}{\sqrt{56}}(\ket{11111222} + perms);\\
&\ket{8}=\frac{1}{\sqrt{56}}(\ket{22222111} + perms);
\,\,\, \ket{9}=\frac{1}{\sqrt{70}}(\ket{11112222} + perms),\\
\end{split}
\end{equation}
where here and in what follows  `$+perms$' means we include all
permutations of tensor indices.

Here the $O'$ invariant polynomial is $x^{8}+y^8+14x^4y^4$, which leads to a VEV proportional to 
\begin{equation}\label{O'vev}
v=[1,1,0,0,0,0,0,0,\frac{14}{\sqrt{70}},0,0,0,0,0,0,0,0,0]
\end{equation}

Where $\ket{1}=\ket{2}$ and $\ket{9}=\frac{14}{\sqrt{70}}\ket{1}$.

$O'$ is a  maximal subgroup of $SU(2)$, so we can be certain our alignment breaks SU(2) uniquely to $O'$.

\subsubsection{$I'$}

Similar to the spontaneous symmetry breaking behavior of the $T'$ case relative to the $A_4$ case with a real $\bf{7}$, the basis for the symmetry breaking to $I'$ with the real \textbf{13} will be the same as for $A_5$ above. Additionally, both groups have the same invariant polynomial so the vacuum directions will be the same.

On the other hand, a complex \textbf{13} arises from the basis of the symmetric tensor with 12 doublet indices:
\begin{equation}
\begin{split}
&\ket{1}=\frac{}{}\ket{111111111111};
\,\,\,  \ket{2}=\frac{}{}\ket{222222222222};
\,\,\, \ket{3}=\frac{1}{\sqrt{12}}(\ket{111111111112} + perms);\\
&\ket{4}=\frac{1}{\sqrt{12}}(\ket{222222222221} + perms);
\,\,\, \ket{5}=\frac{1}{\sqrt{66}}(\ket{111111111122} + perms);\\
&\ket{6}=\frac{1}{\sqrt{66}}(\ket{222222222211} + perms);;
\,\,\, \ket{7}=\frac{1}{\sqrt{220}}(\ket{111111111222} + perms);\\
&\ket{8}=\frac{1}{\sqrt{220}}(\ket{222222222111} + perms);
\,\,\, \ket{9}=\frac{1}{\sqrt{495}}(\ket{111111112222} + perms);\\
&\ket{10}=\frac{1}{\sqrt{495}}(\ket{222222221111} + perms);
\,\,\, \ket{11}=\frac{1}{\sqrt{792}}(\ket{111111122222} + perms);\\
&\ket{12}=\frac{1}{\sqrt{792}}(\ket{222222211111} + perms);
\,\,\, \ket{13}=\frac{1}{\sqrt{924}}(\ket{222222111111} + perms).\\
\end{split}
\end{equation}

Here the $I'$ invariant polynomial is $x^{11}y+11x^6y^6-y^{11}x$, which leads to a VEV proportional to 
\begin{equation}\label{I'vev}
v=[0,0,1,-1,0,0,0,0,0,0,0,0,\sqrt{\frac{11}{12}},0,0,0,0,0,0,0,0,0,0,0,0,0]
\end{equation}

Where clearly $\ket{4}=-\ket{3}$ and $\ket{13}=\sqrt{\frac{11}{12}}\cdot\ket{3}$.

$I'$ is a known maximal subgroup of SU(2), so we can be certain our alignment breaks SU(2) uniquely to $I'$.

\subsection{ Vacuua for $SU(3)$ Potentials}

First let us show that we can get discrete subgroups from continuous groups in a nonminimal way. For this purpose we use the example $SU(3)\rightarrow A_4$ where we break   with a \textbf{15} of $SU(3)$. Then we find vacuua for the minimal cases discussed above. Then finally, for $PSL(2,7)$ we give both a minimal case with a VEV  for the  ${\bf 15'}$ of  $SU(3)$ and a nonminimal breaking via a \textbf{28} of $SU(3)$ using the potential given in eq. (\ref{V28}).

\subsubsection{$A_4$}

The complex 15 dimensional basis needed to break SU(3) to $A_4$ is that of the traceless $3\times 3\times \bar3$ tensor that is symmetric on the first two indices \cite{Luhn:2011ip}.

\begin{equation}
\begin{split}
&\ket{1}=\frac{1}{\sqrt{3}}(\ket{111}-\ket{122}-\ket{212})\\
&\ket{2}=\frac{1}{2\sqrt{6}}(2\cdot\ket{111}+\ket{122}+\ket{212}-3\cdot\ket{133}-3\cdot\ket{313})\\
&\ket{3}=\frac{1}{\sqrt{3}}(\ket{222}-\ket{233}-\ket{323})\\
&\ket{4}=\frac{1}{2\sqrt{6}}(2\cdot\ket{222}+\ket{233}+\ket{323}-3\cdot\ket{211}-3\cdot\ket{121})\\
&\ket{5}=\frac{1}{\sqrt{3}}(\ket{333}-\ket{311}-\ket{131})\\
&\ket{6}=\frac{1}{2\sqrt{6}}(2\cdot\ket{333}+\ket{311}+\ket{131}-3\cdot\ket{322}-3\cdot\ket{232})\\
&\ket{7}=\frac{}{}\ket{112};
\,\,\, \ket{8}=\ket{113};
\,\,\, \ket{9}=\ket{223}\\
&\ket{10}=\frac{}{}\ket{221};
\,\,\, \ket{11}=\ket{331};
\,\,\, \ket{12}=\ket{332}\\
&\ket{13}=\frac{1}{\sqrt{2}}(\ket{123}+\ket{213});\,\,\,
\ket{14}=\frac{1}{\sqrt{2}}(\ket{231}+\ket{321});\,\,\,
\ket{15}=\frac{1}{\sqrt{2}}(\ket{312}+\ket{132})\\
\end{split}
\end{equation}

Because this tensor is symmetric on only two indices we find that the invariant should be of degree 2 in the variables $x,y,z$ and degree 1 in the conjugate variables, $x^*,y^*,z^*$. Inputting the trial polynomial $xyz^*$ into the Reynolds operator produces the invariant: $xyz^*+yzx^*+xzy^*$. In this basis the VEV is proportional to
\begin{equation}\label{A4su3vev}
v=[0,0,0,0,0,0,0,0,0,0,0,0,1,1,1,0,0,0,0,0,0,0,0,0,0,0,0,0,0,0]
\end{equation}
i.e., where $\ket{13}=\ket{14}=\ket{15}$ with all other components zero.

One can examine the generators of $A_4$ and SU(3) to see that this VEV breaks SU(3) uniquely to $A_4$, see \cite{Luhn:2011ip}.

\subsubsection{$T_7$}

The invariant tensor object and therefore our basis for $T_7$ is the same as for $A_4$ above. The invariant polynomial in this case is $x^2y^*+y^2z^*+z^2x^*$ and the corresponding VEV is proportional to
\begin{equation}\label{T7vev}
v=[0,0,0,0,0,0,1,0,1,0,1,0,0,0,0,0,0,0,0,0,0,0,0,0,0,0,0,0,0,0]
\end{equation}
where $\ket{7}=\ket{9}=\ket{11}$.
 
Similarly to the $A_4$ case, one can verify this VEV uniquely breaks $SU(3)$ to $T_7$ by examining how the $T_7$ generators operate on $v$, see \cite{Luhn:2011ip}.

\subsubsection{$\Delta(27)$}

For $\Delta(27)$, the relevant invariant tensor is the fully symmetric part of $3\times 3\times 3$ with 10 independent components 
\begin{equation}
\begin{split}
&\ket{1}=\ket{111};\,\,\,
\ket{2}=\ket{222};\,\,\,
\ket{3}=\ket{333}\\
&\ket{4}=\frac{1}{\sqrt{3}}(\ket{112}+\ket{121}+\ket{211});\,\,\,
\ket{5}=\frac{1}{\sqrt{3}}(\ket{113}+\ket{131}+\ket{311})\\  
&\ket{6}=\frac{1}{\sqrt{3}}(\ket{221}+\ket{212}+\ket{122});\,\,\,
\ket{7}=\frac{1}{\sqrt{3}}(\ket{223}+\ket{232}+\ket{322})\\
&\ket{8}=\frac{1}{\sqrt{3}}(\ket{331}+\ket{313}+\ket{133});\,\,\,
\ket{9}=\frac{1}{\sqrt{3}}(\ket{332}+\ket{323}+\ket{233})\\
&\ket{10}=\frac{1}{\sqrt{6}}(\ket{123}+\ket{231}+\ket{312}+\ket{321}+\ket{213}+\ket{132})\\
\end{split}
\end{equation}

The invariant polynomial is $x^3+y^3+z^3$, which gives us a VEV proportional to
\begin{equation}\label{D27VEV}
v=[1,1,1,0,0,0,0,0,0,0,0,0,0,0,0,0,0,0,0,0]
\end{equation}
Again explicit forms of the generators can be examined in order to verify the uniqueness of this VEV for breaking from 
$SU(3)$ to $\Delta(27)$ \cite{Luhn:2011ip}.

\subsubsection{$PSL(2,7)$}

Our basis for the $\mathbf{15'}$ is that of the fully symmetric $3\times 3\times 3\times 3$ tensor
\begin{equation}
\begin{split}
&\ket{1}=\ket{1111};\,\,\,
\ket{2}=\ket{2222};\,\,\,
\ket{3}=\ket{3333}\\
&\ket{4}=\frac{1}{2}(\ket{1112}+\ket{1121}+\ket{1211}+\ket{2111});\,\,\,
\ket{5}=\frac{1}{2}(\ket{1113}+\ket{1131}+\ket{1311}+\ket{3111})\\  
&\ket{6}=\frac{1}{2}(\ket{2221}+\ket{2212}+\ket{2122}+\ket{1222});\,\,\,
\ket{7}=\frac{1}{2}(\ket{2223}+\ket{2232}+\ket{2322}+\ket{3222})\\
&\ket{8}=\frac{1}{2}(\ket{3331}+\ket{3313}+\ket{3133}+\ket{1333});\,\,\,
\ket{9}=\frac{1}{2}(\ket{3332}+\ket{3323}+\ket{3233}+\ket{2333})\\
&\ket{10}=\frac{1}{\sqrt{6}}(\ket{1122}+perms);\,\,\,
\ket{11}=\frac{1}{\sqrt{6}}(\ket{1133}+perms);\,\,\,
\ket{12}=\frac{1}{\sqrt{6}}(\ket{2233}+perms)\\
&\ket{13}=\frac{1}{\sqrt{12}}(\ket{1123}+perms);\,\,\,
\ket{14}=\frac{1}{\sqrt{12}}(\ket{2213}+perms);\,\,\,
\ket{15}=\frac{1}{\sqrt{12}}(\ket{3312}+perms)\\
\end{split}
\end{equation}

The relevant invariant polynomial  is $x^3z+y^3x+z^3y$ \cite{Merle:2011vy}, which gives a VEV proportional to
\begin{equation}\label{PSLVEV}
v=[0,0,0,0,1,1,0,0,1,0,0,0,0,0,0,0,0,0,0,0,0,0,0,0,0,0,0,0,0,0]
\end{equation}
where the nonvanishing vacuum components are $\ket{5}=\ket{6}=\ket{9}$. We can be sure we have broken to  the correct subgroup {\footnote{Luhn \cite{C.Luhn} has shown that the VEV in eq. (\ref{PSLVEV}) has a  $Z_{28}$ symmetry and the vacuum of the potential $V_{15'}$ in eq. (\ref{V15'}) is also  symmetric under this symmetry. However, other terms in the Lagrangian will violate this $Z_{28}$, e.g., the Yukawa terms. As it is a discrete symmetry, its breaking can not lead to a pseudo Goldstone boson, but there could be other phenomenological consequences of this $Z_{28}$ that would be interesting to explore.}} because $PSL(2,7)$ is known to be a maximal in $SU(3)$.

Finally, for the \textbf{28} of $SU(3)$ we have the basis for the fully symmetric ${\bf 3}^6$ tensor of the form
\begin{equation}
\begin{split}
&\ket{1}=\ket{111111};\,\,\,
\ket{2}=\ket{222222};\,\,\,
\ket{3}=\ket{333333};\,\,\,
\ket{4}=\frac{1}{\sqrt{6}}(\ket{111112}+perms)\\
&\ket{5}=\frac{1}{\sqrt{6}}(\ket{111113}+perms);\,\,\,
\ket{6}=\frac{1}{\sqrt{6}}(\ket{222221}+perms)\\ 
&\ket{7}=\frac{1}{\sqrt{6}}(\ket{222223}+perms);\,\,\,
\ket{8}=\frac{1}{\sqrt{6}}(\ket{333331}+perms)\\ 
&\ket{9}=\frac{1}{\sqrt{6}}(\ket{333332}+perms);\,\,\,
\ket{10}=\frac{1}{\sqrt{15}}(\ket{111122}+perms)\\
&\ket{11}=\frac{1}{\sqrt{15}}(\ket{111133}+perms);\,\,\,
\ket{12}=\frac{1}{\sqrt{15}}(\ket{222211}+perms)\\
&\ket{13}=\frac{1}{\sqrt{15}}(\ket{222233}+perms);\,\,\,
\ket{14}=\frac{1}{\sqrt{15}}(\ket{333311}+perms)\\
&\ket{15}=\frac{1}{\sqrt{15}}(\ket{333322}+perms);\,\,\,
\ket{16}=\frac{1}{\sqrt{30}}(\ket{111123}+perms)\\
&\ket{17}=\frac{1}{\sqrt{30}}(\ket{222231}+perms);\,\,\,
\ket{18}=\frac{1}{\sqrt{30}}(\ket{333312}+perms)\\
&\ket{19}=\frac{1}{\sqrt{20}}(\ket{111222}+perms);\,\,\,
\ket{20}=\frac{1}{\sqrt{20}}(\ket{111333}+perms)\\
&\ket{21}=\frac{1}{\sqrt{20}}(\ket{222333}+perms);\,\,\,
\ket{22}=\frac{1}{\sqrt{60}}(\ket{111223}+perms)\\
&\ket{23}=\frac{1}{\sqrt{60}}(\ket{111332}+perms);\,\,\,
\ket{24}=\frac{1}{\sqrt{60}}(\ket{222113}+perms)\\
&\ket{25}=\frac{1}{\sqrt{60}}(\ket{222331}+perms);\,\,\, 
\ket{26}=\frac{1}{\sqrt{60}}\ket{333112}+perms);\\
&\ket{27}=\frac{1}{\sqrt{60}}(\ket{333221}+perms)\,\,\,
\ket{28}=\frac{1}{\sqrt{90}}(\ket{112233}+perms)\\
\end{split}
\end{equation}

The necessary invariant polynomial is $x^5y + y^5z + z^5x- 5x^2y^2z^2 $\cite{Merle:2011vy}, which gives real components with VEV proportional to
\begin{equation}\label{28PSLVEV}
v=\left[0,0,0,1,0,0,1,1,0,0,0,0,0,0,0,0,0,0,0,0,0,0,0,0,0,0,0,-\sqrt{\frac{5}{3}}\right]
\end{equation}
i.e., where $\ket{4}=\ket{7}=\ket{8}$, and $\ket{28}=-\sqrt{\frac{5}{3}}\cdot\ket{4}$ and where we recall that  all conjugate components (29--56) are set to zero.

\section{ Vacuum expectation values and Mass Spectra}\label{scalesspectra}

Thus far, we have discussed how to set up potentials corresponding to specific gauge group representations and then found vacuum alignments that can be used to break the gauge symmetry to desired subgroups. In this section we minimize the scalar potentials and show where symmetry breaking in the desired directions are allowed.  We will find the scale of the symmetry breaking and resulting tree level scalar mass states in terms of the coupling constants of the potential. As usual, the minimization conditions of the potential will lead to constraints on the values of these constants. 

\subsection{$SO(3)$ Cases}

\subsubsection{$A_4$ scalar spectrum}

We found earlier that a VEV in the direction \eqref{eq:so3a4vev} will break $SO(3)$ to $A_4$. The actual VEV is proportional to this direction vector, with the constant of proportionality being the scale of the breaking. To determine this scale one must minimize the potential \eqref{eq:V7}. To achieve this we compute the first derivative with respect to each basis state, insert the alignment from \eqref{eq:so3a4vev}, and set this equal to zero. This alignment (and all of our alignments below) will give an equation in terms of one basis state (or one linear combination of basis states). For the present case we solve for $\ket{7}$ and take the positive solution to obtain the VEV
\begin{equation}
{\cal V}=\sqrt{\frac{3m^2}{2(3\lambda+\kappa)}}\,[0,0,0,0,0,0,1]
\end{equation}

As for any non-trivial stable vacuum, $m^2$ must be positive. To have a real value for our breaking scale $3\lambda +\kappa$ must also be positive. We find the scalar mass states by calculating the matrix of second derivatives (the Hessian), inserting the VEV from above, and computing the eigenvalues of the matrix. The resulting values and their multiplicities are given in Table \ref{tab:A4/7}.

\begin{table}[h]
\begin{center}
\begin{tabular}{c|c} 
Value & Multiplicity \\ [1ex] 
\hline\hline
0 & 3\\[2ex]
$4m^2$ & 1\\[2ex]
$\frac{8m^2\kappa}{5(3\lambda+\kappa)}$ & 3\\[2ex]
\end{tabular}
\caption{Scalar mass eigenstates for the SSB pattern $SO(3) \rightarrow A_4$ using a real \textbf{7} of $SO(3)$.}
\label{tab:A4/7}
\end{center}
\end{table}

Looking at Table \ref{tab:4} in the Appendix, we see that the multiplicities of the eigenvalues match up with the branching of the {\textbf{7}} for $SO(3) \rightarrow A_4$, as expected. We  see that there are three zero eigenvalues as expected corresponding to the three Goldstone boson from the breaking of all the generators of $SO(3)$. Constraints on the coupling constants arise from the requirement that at a minimum of the potential, the eigenvalues must all positive or zero. Since $m^2$  and $3\lambda+\kappa$ must be positive,  requiring the third eigenvalue to be positive leads to the constraint $\kappa > 0$ in this case.

\subsubsection{$S_4$}

For $S_4$ we minimize the potential from \eqref{eq:V9} using the alignment \eqref{vevS4}. We obtain a VEV
\begin{equation}
{\cal V}=\sqrt{\frac{25m^2}{4(90\lambda+10\kappa+7\rho+2\tau)}}\,[\sqrt{\frac{7}{5}},0,0,0,0,0,0,1,0]
\end{equation}

A real value for our breaking scale requires $90\lambda+10\kappa+7\rho+2\tau > 0$. 

The scalar mass states are found in Table \ref{tab:S4/9}
\footnote{In order to normalize the eigenvalues for $S_4$ to those in other cases when we use the spherical harmonic basis, we have multiplied all quadratic terms by a factor of $\frac{1}{8}$ and quartic terms by a factor of $\frac{1}{64}$.}
 and are all non-negative if \begin{center}
$5\kappa+8\rho-2\tau>0.$ 
\end{center} 
\begin{table}[h]
\begin{center}
\begin{tabular}{c|c} 
Value & Multiplicity \\ [1ex] 
\hline\hline
0 & 3\\[2ex]
$4m^2$ & 1\\[2ex]
$\frac{5m^2(5\kappa+8\rho-2\tau)}{7(90\lambda+10\kappa+7\rho+2\tau)}$ & 3\\[2ex]
$\frac{20m^2(5\kappa+8\rho-2\tau)}{7(90\lambda+10\kappa+7\rho+2\tau)}$ & 2\\[2ex]
\end{tabular}
\end{center}
\caption{Scalar mass eigenstates for the SSB pattern $SO(3) \rightarrow S_4$ using a real \textbf{9} of $SO(3)$.}
\label{tab:S4/9}
\end{table}
The three zeros correspond to the broken $SO(3)$ generators.

\subsubsection{$A_5$ scalar spectrum}

\vspace{2ex}

For $A_5$, we minimize the potential from \eqref{SO3_13} using the alignment \eqref{vevA5}. We obtain a VEV
\begin{equation}
{\cal V}=\sqrt{\frac{1155m^2}{128(\lambda+140\kappa+84\rho+65\tau+14\nu+9\sigma-2\chi)}}\,[1,0,0,-\sqrt{\frac{21}{2}},0,0,0,,0,0,0,\sqrt{\frac{105}{22}},0]
\end{equation}

A real value for our breaking scale requires $420\lambda+140\kappa+84\rho+65\tau+14\nu+9\sigma-2\chi > 0$. For the scalar mass states
\footnote{Again by expressing our states in terms of spherical harmonics, we obtain different normalizations for our basis states which lead to a different normalization scale for the VEV scale and scalar mass states. To correct this for $A_5$ we have multiplied the quadratic term  by a factor of $\frac{5}{352}$ and the quartic terms by $(\frac{5}{352})^2$  so that our states are now normalized the same way as our other breakings.}
are given in Table \ref{tab:A5/13}.

\begin{table}[h]
\begin{center}
\begin{tabular}{c|c} 
Value & Multiplicity \\ [1ex] 
\hline\hline
0 & 3\\[2ex]
$4m^2$ & 1\\[2ex]
$\frac{28m^2(105\kappa+196\rho+240\tau-14\nu-19\sigma+12\chi)}{33(420\lambda+140\kappa+84\rho+65\tau+14\nu+9\sigma-2\chi}$ & 5\\[2ex]
$\frac{28m^2(14\rho+45\tau+14\nu-11\sigma+18\chi)}{33(420\lambda+140\kappa+84\rho+65\tau+14\nu+9\sigma-2\chi}$ & 4\\[2ex]
\end{tabular}
\end{center}
\caption{Scalar mass eigenstates for the SSB pattern $SO(3) \rightarrow A_5$ using a real \textbf{13} of $SO(3)$.}
\label{tab:A5/13}
\end{table}

We see we have the three zeros corresponding to the broken $SO(3)$ generators and must satisfy the  constraints
\begin{center}
$105\kappa+196\rho+240\tau-14\nu-19\sigma+12\chi>0$ \\[2ex] $14\rho+45\tau+14\nu-11\sigma+18\chi>0.$ \\[2ex]
\end{center}

\subsection{   $SU(2)$ Cases}

\subsubsection{$Q_6$ scalar spectrum}

We  break the symmetry of the potential given in Eq.\eqref{complex7}   with the alignment in Eq.\eqref{Q6vev} to obtain a VEV
\begin{equation}
{\cal V}=\sqrt{\frac{m^2}{2(2\lambda+\kappa+\rho+\tau)}}\,[1,1,0,0,0,0,0,0,0,0,0,0,0,0]
\end{equation}

Thus we require $\kappa+2\lambda+\rho+\tau>0$. The eigenvalues of the Hessian are given in Table \ref{tab:Q6/c7}.
\begin{table}[h]
\begin{center}
\begin{tabular}{c|c} 
Value & Multiplicity \\ [1ex] 
 \hline\hline
0 & 4\\[2ex]
$4m^2$ & 1\\[2ex]
$\frac{2m^2\kappa}{3(2\lambda+\kappa+\rho+\tau)}$ & 2\\[2ex]
$\frac{4m^2(\kappa+\rho+\tau)}{2\lambda+\kappa+\rho+\tau}$ & 1\\[2ex]
$\frac{-3m^2(2\rho+3\tau)}{5(2\lambda+\kappa+\rho+\tau)}$ & 1\\[2ex]
$\frac{-2m^2(2\rho+3\tau)}{5(2\lambda+\kappa+\rho+\tau)}$ & 2\\[2ex]
$\frac{-m^2(6\rho+7\tau)}{5(2\lambda+\kappa+\rho+\tau)}$ & 1\\[2ex]
$\frac{-2m^2(8\rho+9\tau)}{15(2\lambda+\kappa+\rho+\tau)}$ & 2\\[2ex]
\end{tabular}
\end{center}
\caption{Scalar mass eigenstates for the SSB pattern $SU(2) \rightarrow Q_6$ using a complex \textbf{7} of $SU(2)$.}
\label{tab:Q6/c7}
\end{table}

The constraints from these mass eigenvalues are
\begin{center}
$\kappa>0$ \\[2ex] $\kappa>-(\rho+\tau)$ \\[2ex] $2\rho+3\tau,\:6\rho+7\tau,\:8\rho+9\tau<0$
\end{center}  
There are clearly stable minima when $\lambda>0$, $\kappa>0$, $\rho<0$ and $\tau<0$.
The extra zero eigenvalue comes from breaking  an accidental $U(1)$ phase symmetry.
This gives rise to a pseudo-goldstone boson that can gain a mass through quantum corrections.

\subsubsection{$T'$  scalar spectrum}

The potential and the vacuum alignment of the breaking of  of SU(2) to T$'$ with a real \textbf{7}  are the same as for SO(3)$\rightarrow A_4$. Therefore the breaking scale and the mass states will be exactly the same, as the two models can only be differentiated by the non-scalar part of the Lagrangian. 

For a the breaking with a complex \textbf{7} we minimize the potential in Eq.\eqref{complex7} but this time using the alignment Eq.\eqref{vevTprime} to obtain the VEV
\begin{equation}
{\cal V}=\sqrt{\frac{3m^2}{(12\lambda+6\kappa+4\rho+3\tau)}}\,[0,-1,0,0,0,1,0,0,0,0,0,0,0,0]
\end{equation}
which leads to the constraint that $12\lambda+6\kappa+4\rho+3\tau>0$. The eigenvalues of the Hessian are shown in Table \ref{tab:T'/c7}.
\begin{table}[h]
\begin{center}
\begin{tabular}{c|c} 
Value & Multiplicity \\ [1ex] 
 \hline\hline
0 & 4\\[2ex]
$4m^2$ & 1\\[2ex]
$\frac{12m^2\tau}{5(12\lambda+6\kappa+4\rho+3\tau)}$ & 3\\[2ex]
$\frac{16m^2(2\rho+3\tau)}{5(12\lambda+6\kappa+4\rho+3\tau)}$ & 3\\[2ex]
$\frac{4m^2(8\kappa+8\rho+9\tau)}{3(12\lambda+6\kappa+4\rho+3\tau)}$ & 3\\[2ex]
\end{tabular}
\end{center}
\caption{Scalar mass eigenstates for the SSB pattern $SU(2) \rightarrow T'$ using a complex \textbf{7} of $SU(2)$.}
\label{tab:T'/c7}
\end{table}

From the requirement of positive eigenvalues we deduce the constraints
\begin{center}
$\tau>0$ \\[2ex] $\rho>-\frac{3}{2}\tau$ \\[2ex] $\frac{3}{8}\tau>\kappa>-8\rho-\frac{9}{8}\tau$
\end{center}

As in the $Q_6$ example, the extra zero eigenvalue is a result of breaking the accidental $U(1)$ phase symmetry in the potential. 
\subsubsection{$O'$ scalar spectrum}

The breaking scale and scalar mass spectrum of $SU(2)$ to $O'$ with a real $\bf 9$ is exactly the same as that for $SO(3)$ to $S_4$, where differences between  two models would come from the non-scalar part of the Lagrangian. 

For a complex \textbf{9} we minimize the potential in Eq.\eqref{complex9} using the alignment Eq.\eqref{O'vev} and obtain a VEV
\begin{equation}
\begin{split}
{\cal V}=\sqrt{\frac{25m^2}{4(60\lambda+30\kappa+20\rho+15\tau+14\sigma)}}\\
\times[1,1,0,0,0,0,0,0,\frac{14}{\sqrt{70}},0,0,0,0,0,0,0,0,0]
\end{split}
\end{equation}

Thus $60\lambda+30\kappa+20\rho+15\tau+14\sigma$ must be $> 0$. The eigenvalues of the Hessian (see Table \ref{tab:O'/c9}) are  all real and positive semidefinite for positive scalar quartic couplings, while more detailed constraints on the scalar quartics can clearly be extracted from the individual mass eigenvalues. There are 3 zeros corresponding to the 3 broken $SU(2)$ generators, as well as an extra zero from breaking the $U(1)$ phase symmetry. 

\begin{table}[h]
\begin{center}
\begin{tabular}{c|c} 
Value & Multiplicity \\ [1ex] 
\hline\hline
0 & 4 \\ [2ex]
$4m^2$ & 1 \\[2ex]
$\frac{-24m^2\sigma}{7(60\lambda+30\kappa+20\rho+15\tau+14\sigma)}$ & 2 \\[2ex]
$\frac{5m^2(10\rho+15\tau+16\sigma)}{7(60\lambda+30\kappa+20\rho+15\tau+14\sigma)}$ & 3 \\[2ex]
$\frac{20m^2(10\rho+15\tau+16\sigma)}{7(60\lambda+30\kappa+20\rho+15\tau+14\sigma)}$ & 2 \\[2ex]
$\frac{2m^2(25\kappa+25\rho+25\tau+24\sigma)}{60\lambda+30\kappa+20\rho+15\tau+14\sigma}$ & 3 \\[2ex]
$\frac{3m^2(25\tau+32\sigma)}{7(60\lambda+30\kappa+20\rho+15\tau+14\sigma)}$ & 3 \\[2ex]
\end{tabular}
\end{center}
\caption{Scalar mass eigenstates for the SSB pattern $SU(2) \rightarrow O'$ using a complex \textbf{9} of $SU(2)$.}
\label{tab:O'/c9}
\end{table}
We have the additional constraints
\begin{center}
$\sigma<0$ \\[2ex] $10\rho+15\tau+16\sigma>0$ \\[2ex] $25\kappa+25\rho+25\tau+24\sigma>0$\\[2ex]
and\\[2ex]$25\tau+32\sigma>0.$
\end{center}

\subsubsection{$I'$ scalar spectrum}

The breaking of $SU(2)$ to $I'$ and  $SO(3)$ to $A_5$ with a real \textbf{13}, are completely analogous to the breakings of $SU(2)$ and $SO(3)$ to  $T'$ and $A_4$ respectively with a real \textbf{7}.

For a complex \textbf{13} we minimize the potential of Eq.\eqref{V13*} using the alignment Eq.\eqref{I'vev} and obtain a VEV
\begin{equation}
\begin{split}
{\cal V}=7\sqrt{\frac{6m^2}{5(420\lambda+210\kappa+140\rho+105\tau+84\nu+70\sigma+65\chi)}}\\
\times[0,0,1,-1,0,0,0,0,0,0,0,0,\sqrt{\frac{11}{7}},0,0,0,0,0,0,0,0,0,0,0,0,0]
\end{split}
\end{equation}

Thus $420\lambda+210\kappa+140\rho+105\tau+84\nu+70\sigma+65\chi$ must be $> 0$. The eigenvalues of the Hessian (see Table \ref{tab:I'/c13}) are  all real and positive semidefinite for positive scalar quartic couplings. (More detailed constraints on the scalar quartics can clearly be extracted from the individual mass eigenvalues.)
\begin{table}[h]
\begin{center}
 \begin{tabular}{c|c} 
 Value & Multiplicity \\ [1ex] 
 \hline\hline
 0 & 4 \\ [2ex]
 $4m^2$ & 1 \\[2ex]
$\frac{28m^2(14\nu+35\sigma+45\chi)}{33(420\lambda+210\kappa+140\rho+105\tau+84\nu+70\sigma+65\chi)}$ & 4 \\[2ex]
$\frac{5m^2(49\sigma+72\chi)}{33(420\lambda+210\kappa+140\rho+105\tau+84\nu+70\sigma+65\chi)}$ & 5 \\[2ex]
$\frac{14m^2(210\rho+315\tau+392\nu+455\sigma+480\chi)}{33(420\lambda+210\kappa+140\rho+105\tau+84\nu+70\sigma+65\chi)}$ & 5 \\[2ex]
$\frac{m^2(980\kappa+980\rho+882\tau+784\nu+735\sigma+720\chi)}{3(420\lambda+210\kappa+140\rho+105\tau+84\nu+70\sigma+65\chi)}$ & 3 \\[2ex]
$\frac{4m^2(441\tau+882\nu+1225\sigma+1350\chi)}{33(420\lambda+210\kappa+140\rho+105\tau+84\nu+70\sigma+65\chi)}$ & 4 \\[2ex] 
\end{tabular}
\end{center}
\caption{Scalar mass eigenstates for the SSB pattern $SU(2) \rightarrow I'$ using a complex \textbf{13} of $SU(2)$.}
\label{tab:I'/c13}
\end{table}
There are 3 zeros corresponding to the 3 broken $SU(2)$ generators, as well as an extra zero from breaking the $U(1)$ phase symmetry.

\subsection{   SU(3) cases}
\subsubsection{$A_4$ scalar spectrum}

For the nonminimal breaking $SU(3) \rightarrow A_4$ we minimize the potential Eq.\eqref{V15} and use the alignment Eq.\eqref{A4su3vev} to get the VEV \cite{Luhn:2011ip}
\begin{equation}
{\cal V}=\sqrt{\frac{m^2}{2(3\lambda+\eta+\kappa +\rho+\tau)}}\,[0,0,0,0,0,0,0,0,0,0,0,0,1,1,1,0,0,0,0,0,0,0,0,0,0,0,0,0,0,0]
\end{equation}

Thus $3\lambda+\eta+\kappa+\rho+\tau$ must be $> 0$. The eigenvalues of the Hessian are are shown in Table \ref{tab:A4/15ofSU(3)}.
\begin{table}[h]
 \begin{center}
 \begin{tabular}{c|c} 
 Value & Multiplicity \\ [1ex] 
 \hline\hline
 0 & 9 \\ [2ex]
 $4m^2$ & 1 \\[2ex]
 $\frac{m^2(-2\eta+\kappa-2\rho+4\tau)}{3\lambda+\eta+\kappa+ \rho+\tau}$ & 2 \\[2ex]
 $\frac{-3m^2\eta}{3\lambda+\eta+\kappa+ \rho+\tau}$ & 6 \\[2ex]
 $\frac{m^2}{4(3\lambda+\eta+\kappa+ \rho+\tau)}(5\kappa +2\rho+4\tau+\sqrt{(4\tau+2\rho-3\kappa)^2+16(\rho+\kappa+2\eta)^2})$ & 3 \\[2ex]
 $\frac{m^2}{4(3\lambda+\eta+\kappa +\rho+\tau)}(5\kappa +2\rho+4\tau-\sqrt{(4\tau+2\rho-3\kappa)^2+16(\rho+\kappa+2\eta)^2})$ & 3 \\[2ex]
 $\frac{m^2}{2(3\lambda+\eta+\kappa+ \rho+\tau)}(3\kappa -5\eta-2\rho+4\tau+\frac{1}{3}\sqrt{(9\eta-7\kappa+10\rho-4\tau)^2+8(\rho+2\kappa-4\tau)^2})$ & 3 \\[2ex]
 $\frac{m^2}{2(3\lambda+\eta+\kappa+ \rho+\tau)}(3\kappa -5\eta-2\rho+4\tau-\frac{1}{3}\sqrt{(9\eta-7\kappa+10\rho-4\tau)^2+8(\rho+2\kappa-4\tau)^2})$ & 3 \\  [2ex]
\end{tabular}
\end{center}
\caption{Scalar mass eigenstates for the SSB pattern $SU(3) \rightarrow A_4$ using a  \textbf{15} of $SU(3)$.}
\label{tab:A4/15ofSU(3)}
\end{table}

We expect eight zeros corresponding to the broken generators of $SU(3)$, but again an extra zero eigenvalue arises from breaking the accidental $U(1)$ phase symmetry. 
%This can not be remedied by including cubic terms in the potential because the VEV happens to be unaffected by the addition of these terms.   
As for constraints, we can readily see that
\begin{center}
$\eta<0,$ \\[2ex] 
$5\kappa +2\rho+4\tau>\sqrt{(4\tau+2\rho-3\kappa)^2+16(\rho+\kappa+2\eta)^2},$ \\[2ex] 
\end{center}  
and
\begin{center}
$3\kappa -5\eta-2\rho+4\tau>\frac{1}{3}\sqrt{(9\eta-7\kappa+10\rho-4\tau)^2+8(\rho+2\kappa-4\tau)^2}$
\end{center}  
are required. An example of where all these constraints can be satisfied is
\begin{center}
$2\rho = 3 \kappa$, $\rho +\kappa =-2|\eta|$, and $5 \kappa+3\rho>0$, where $ \kappa,\rho,$ and $ \tau>0.$
\end{center}

\subsubsection{$T_7$ scalar spectrum}

For this breaking we again minimize Eq.\eqref{V15}, now using the alignment Eq.\eqref{T7vev} to obtain the VEV \cite{Luhn:2011ip}
\begin{equation}
{\cal V}=\sqrt{\frac{m^2}{2(3\lambda+\kappa+ \rho+\tau)}}\,[0,0,0,0,0,0,0,0,0,0,0,0,1,1,1,0,0,0,0,0,0,0,0,0,0,0,0,0,0,0]
\end{equation}

Thus $3\lambda+\kappa+\rho+\tau$ must be $> 0$. The eigenvalues of the Hessian are shown in Table \ref{tab:T7/15},
\begin{table}[h]
\begin{center}
\begin{tabular}{c|c} 
Value & Multiplicity \\ [1ex] 
\hline\hline
0 & 9 \\[2ex]
$4m^2$ & 1 \\[2ex]
$\frac{2(2\kappa-\rho+2\tau)m^2}{\kappa + 3\lambda +\rho +\tau}$ & 2\\[2ex]
$\frac{m^2}{12(3\lambda+\kappa+\rho+\tau)}\times \, \alpha$ & 6\\[2ex]
$\frac{m^2}{12(3\lambda+\kappa +\rho+\tau)}\times \, \beta$  & 6\\[2ex]
$\frac{m^2}{12(3\lambda+\kappa +\rho+\tau)}\times \, \gamma$ & 6\\[2ex]
\end{tabular}
\end{center}
\caption{Scalar mass eigenstates for the SSB pattern $SU(3) \rightarrow T_7$ using a  \textbf{15} of $SU(3)$.}
\label{tab:T7/15}
\end{table}
where $\alpha,\,\beta,\,\gamma$ are the three roots of the polynomial $10368\eta^2(\rho-\kappa-\tau) +3888\eta\rho^2-15552\eta\kappa\tau+(648\eta^2+972\eta\kappa-648\eta\rho-180\rho^2+1296\eta\tau+720\kappa\tau)x+(6\rho-54\eta-21\kappa-36\tau)x^2+x^3$. 
We have the constraints
\begin{center}
$2\kappa-\rho+2\tau>0$\\[2ex]and\\[2ex]$\alpha,\,\beta,\,\gamma>0.$
\end{center}

The extra zero is once again due to breaking an accidental $U(1)$ symmetry. We cannot remedy this by including cubic terms this time, because we need the couplings on those terms to vanish in order to have a stable minimum. 
Numerical studies show that there is a range of scalar quartic coupling constant values where the minimum is stable. An example of such numerical analysis will be discussed below.

\subsubsection{$\Delta(27)$ scalar spectrum}

Minimizing the potential of Eq.\eqref{V10} with the alignment Eq.\eqref{D27VEV} we obtain a VEV
\begin{equation}
{\cal V}=\sqrt{\frac{m^2}{2(3\lambda+\kappa)}}\,[1,1,1,0,0,0,0,0,0,0,0,0,0,0,0,0,0,0,0,0]
\end{equation}
giving the constraint $3\lambda+\kappa>0$. The eigenvalues of the Hessian are in Table \ref{tab:Delta27/10},
\begin{table}[h]
\begin{center}
\begin{tabular}{c|c} 
Value & Multiplicity \\ [1ex] 
\hline\hline
0 & 11 \\[2ex]
$4m^2$ & 1 \\[2ex]
$\frac{4\kappa m^2}{3\lambda+\kappa}$ & 2\\[2ex]
$\frac{\kappa m^2}{3(3\lambda+\kappa)}$ & 6\\[2ex]
\end{tabular}
\end{center}
\caption{Scalar mass eigenstates for the SSB pattern $SU(3) \rightarrow \Delta(27)$ using a  \textbf{10} of $SU(3)$.}
\label{tab:Delta27/10}
\end{table}
from which we see that $\kappa>0$ is required. Again we have an extra zero from an accidental $U(1)$, and in this case the cubic terms vanish upon summation. The two extra zeros are the result of an additional $\Delta(27)$ singlet within the \textbf{10}. (For a detailled explanation see \cite{Luhn:2011ip}.)

\subsubsection{$PSL(2,7)$ scalar spectrum}
%Still need to address Luhns VEV symmetry
Minimizing the potential of Eq.\eqref{V15'} with the alignment Eq.\eqref{PSLVEV} we obtain a VEV
\begin{equation}
{\cal V}=\sqrt{\frac{m^2}{6\lambda+2\kappa+\rho}}\,[0,0,0,0,1,1,0,0,1,0,0,0,0,0,0,0,0,0,0,0,0,0,0,0,0,0,0,0,0,0]
\end{equation}

Thus $6\lambda+2\kappa+ \rho>0$. The eigenvalues of the Hessian are found in Table \ref{tab:PSL27/15'}
\begin{table}[h]
\begin{center}
\begin{tabular}{c|c} 
Value & Multiplicity \\ [1ex] 
\hline\hline
0 & 9 \\[2ex]
$4m^2$ & 1 \\[2ex]
$\frac{(7\kappa+8\rho)m^2}{2(6\lambda+2\kappa +\rho)}$ & 8\\[2ex]
$\frac{2(3-\sqrt{2})\rho m^2}{3(6\lambda+2\kappa +\rho)}$ & 6\\[2ex]
$\frac{2(3+\sqrt{2})\rho m^2}{3(6\lambda+2\kappa +\rho)}$  & 6\\[2ex]
\end{tabular}
\end{center}
\caption{Scalar mass eigenstates for the SSB pattern $SU(3) \rightarrow PSL(2,7)$ using a  ${\bf 15'}$ of $SU(3)$.}
\label{tab:PSL27/15'}
\end{table}
from which we get the constraints
\begin{center}
$\rho>0$ \hspace{.5cm} and  \hspace{.5cm}  $7\kappa+8\rho>0$.
\end{center}

We once again have an extra zero, but this time it is possible to include cubic terms to break the $ U(1)$ phase. The two cubic terms we can include are
\begin{equation}
\begin{split}
  \epsilon_{imq}\epsilon_{jnr}&\epsilon_{lpt}T^{ijkl}T^{mnop}T^{qrst}\\
&\text{and}\\
  \epsilon^{imq}\epsilon^{jnr}&\epsilon^{lpt}T_{ijkl}T_{mnop}T_{qrst}\\
\end{split}
\end{equation}
which are Hermitian conjugates and are included in the potential with the same real coupling constant, $\zeta$. The VEV scale for the potential including the cubic is now
\begin{equation}
\frac{-3\zeta \pm \sqrt{9\zeta^2+4m^2(6\lambda+2\kappa+\rho)}}{2(6\lambda+2\kappa+\rho)}
\end{equation}

Notice that there may be two possible solutions. The constraint that must hold in both cases is $9\zeta^2 +4m^2(2\kappa+6\lambda+\rho)\geq 0$.

Calculating the eigenvalues of the Hessian produces solutions involving the roots of very large polynomial which is much too large to display, but it is notable that it does produce 8 zeros rather than 9. Furthermore, following the usual procedure, but this time numerically where for simplicity setting all quartic coupling constants to unity, the quadratic coupling to -1 and the cubic to .001 (these values are selected to ensure a stable minimum) produces a VEV scale of approximately 0.3335 and eigenvalues 
\begin{table}[h]
\begin{center}
\begin{tabular}{c|c} 
Value & Multiplicity \\ [1ex] 
\hline\hline
0 & 8 \\[2ex]
$4.002$ & 1 \\[2ex]
$.006003$ & 1\\[2ex]
$0.323647$ & 6\\[2ex]
$0.125244$ & 6\\[2ex]
$0.838169$ & 8\\[2ex]
\end{tabular}
\end{center}
\caption{Numerical results where cubic terms are included  for the scalar mass eigenstates of the SSB pattern $SU(3) \rightarrow PSL(2,7)$ using a  ${\bf 15'}$   of $SU(3)$.}
\label{tab:num15'}
\end{table}
whose multiplicities match the branching rules SU(3)$\rightarrow PSL(2,7)$,
as shown in Table \ref{tab:num15'}.
The degeneracy of the pseudo-Goldstone mass with that of the true Goldstones is lifted as expected, as can be seen in Table \ref{tab:num15'}. Finally note that although we have set all the coupling constants except for the cubic to integer values, we can easily rescale them 
 to smaller values to be sure we are in the perturbative regime of the theory without disturbing the stability of the result. Specifically, while the scalar quartic couplings in the numerical example are not in the perturbative range, we can rescale all the quartics by a factor $s$ and $\zeta$ by a factor $\sqrt{s}$. This leaves the eigenvalues unchanged and puts us into the perturbative regime.
%\vspace{10ex}

Moving on to the \textbf{28}, we minimize the potential in Eq.\eqref{V28} with the alignment Eq.\eqref{28PSLVEV} to obtain a VEV
\begin{equation}
{\cal V}=\sqrt{\frac{9m^2}{2(42\lambda+14\kappa+7\rho+6\tau)}}
[0,0,0,1,0,0,1,1,0,0,0,0,0,0,0,0,0,0,0,0,0,0,0,0,0,0,0,-\sqrt{\frac{5}{3}},...]
\end{equation}
 Thus $42\lambda+14\kappa+7\rho+6\tau>0$. The eigenvalues of the Hessian are given in Table \ref{tab:28}
\begin{table}[h]
\begin{center}
\begin{tabular}{c|c} 
Value & Multiplicity \\ [1ex] 
\hline\hline
0 & 16 \\[2ex]
$4m^2$ & 1 \\[2ex]
$\frac{4(7\rho+9\tau)m^2}{5(42\lambda+14\kappa+7\rho+6\tau)}$ & 7\\[2ex]
$\frac{(21\kappa+20\rho+18\tau)m^2}{42\lambda+14\kappa+7\rho+6\tau}$ & 8\\[2ex]
$\frac{1}{200(42\lambda+14\kappa+7\rho+6\tau)^2}\times A$  & 6\\[2ex]
$\frac{1}{200(42\lambda+14\kappa+7\rho+6\tau)^2}\times B$  & 6\\[2ex]
$\frac{1}{200(42\lambda+14\kappa+7\rho+6\tau)^2}\times C$  & 6\\[2ex]
$\frac{1}{200(42\lambda+14\kappa+7\rho+6\tau)^2}\times D$  & 6\\[2ex]
\end{tabular}
\end{center}
\caption{Scalar mass eigenstates for the SSB pattern $SU(3) \rightarrow PSL(2,7)$ using a  {\textbf {28}} of $SU(3)$.}
\label{tab:28}
\end{table}
Further  constraints are
\begin{center}
$7\rho+9\tau>0$\\[2ex]$21\kappa+20\rho+18\tau>0$\\[2ex]$A,\,B,\,C,\,D>0$
\end{center}
where A, B, C, and D are the roots of a very large quartic polynomial. Numerical work shows that all four roots can be positive, simultaneously leading to all positive eigenvalues in Table \ref{tab:28} and a stable minimum when the other constraints are also satisfied. We see that there are eight zeros from the broken generators of SU(3), and one zero from breaking the broken U(1) phase. But unique to this breaking we have seven extra zeros, which  implies that there are seven more broken generators from an accidental symmetry of the Lagrangian that we have so far been unable to identify, leading to a total of 8 pseudo-Goldstone bosons. 

\subsection{Symmetry Breaking Summary}

Let us briefly summarize our results. We have shown that we can break from $G$ to $\Gamma $ for the gauge and discrete groups listed in the introduction. The minima can be stable since none of the eigenvalues of the scalars are negative for allowed regions of parameter space. Zero eigenvalues correspond to Goldstone bosons in each case and to additional pseudo-Goldstone bosons in several cases. Specifically for the cases we have studied of $SO(3)$ breaking to a discrete symmetry the results are summarized in Table \ref{tab:SO(3)sum}. The $G$ subscript indicates the Goldstones.
In each case the masses of the particles in different discrete group irreps are all different, so the initial degeneracy of the scalar masses is lifted to the extent allowed by the discrete group. 
\begin{table}[h]
\begin{center}
\begin{tabular}{c|c} 
SSB pattern & decomposition \\ [1ex] 
\hline\hline
$SO(3) \rightarrow A_4$ & ${\bf 7 \rightarrow 1 + 3_{_G}+ 3}$\\[2ex]
$SO(3) \rightarrow S_4$& ${\bf 9 \rightarrow 1 + 2+3_{_G}+ 3}$\\[2ex]
$SO(3) \rightarrow A_5$& ${\bf 13 \rightarrow 1 + 3_{_G}+ 4+5}$\\[2ex]
\end{tabular}
\caption{Scalar mass eigenstates for the SSB patterns $SO(3) \rightarrow A_4,$ $S_4$ and $A_5$ using  real irreps of $SO(3)$.}
\label{tab:SO(3)sum}
\end{center}
\end{table}
For the  cases of $SU(2)$ breaking to discrete symmetries, the results are summarized in Table \ref{tab:SU(2)sum}. Again all the discrete group irreps correspond to different masses except for the zero eigenvalue states where we have indicated the true Goldstones and the pseudo-Goldstones (by subscripts pGB) due to breaking of the phase symmetry on the potentials. The subscript $c$ indicates that the irreps are complexified and the decompositions are written in terms of real components. The results begin to become more complicated for the $SU(3)$ cases we have investigated, and this can be seen in Table \ref{tab:SU(3)sum}. Now some irreps masses have become degenerate and we have indicated these cases by collecting those discrete group irreps with parentheses and labeling the collection with a $deg.$ subscript. All the cases have a pseudo-Goldstones associated with breaking of phase invariance. The breaking to $T_7$ with a ${\bf 10}$ leads to two additional pseudo-Goldstones as discussed in \cite{Luhn:2011ip} and the breaking to $PSL(2,7)$ with a ${\bf 28}$  has seven additional pseudo-Goldstones. Since the ${\bf 28}$ was derived from ${\bf 3^6}$ one could conjecture that the potential has a $Spin(6)\sim SU(4)$
accidental symmetry that contains the gauged $SU(3)$, and that the VEV breaks all 15 $SU(4)$ plus the phase to give a total of 16 massless states.
Finally, recall that for  the breaking to $PSL(2,7)$ with a ${\bf 15'}$ we have shown that phase symmetry can be avoided if we add  cubic terms, hence there is no pseudo-Goldstone after SSB in that case, see Table \ref{tab:num15'}.
\begin{table}[h]
\begin{center}
\begin{tabular}{c|c} 
SSB pattern & decomposition \\ [1ex] 
\hline\hline
$SU(2) \rightarrow Q_6$ & ${\bf 7_c \rightarrow 1 + 1'+(1'+ 2')_{_G} +2'+1_{_{pGB}} + 1'+1'+2'+2'}$\\[2ex]
$SU(2) \rightarrow T'$& ${\bf 7_c \rightarrow 1 + 3_{_G}+ 3+1_{_{pGB}} + 3 + 3}$\\[2ex]
$SU(2) \rightarrow O'$& ${\bf 9_c \rightarrow 1 +2+ 3_{_G}+ 3+1_{_{pGB}} +2+ 3 + 3}$\\[2ex]
$SU(2) \rightarrow I'$& ${\bf 13_c \rightarrow 1 + 3_{_G}+ 4+5+1_{_{pGB}} + 3 + 4+5}$\\[2ex]
\end{tabular}
\caption{Scalar mass eigenstates for the SSB patterns $SU(2) \rightarrow Q_6,$ $T'$, $O'$ and $I'$ using complexified irreps of $SO(3)$.}
\label{tab:SU(2)sum}
\end{center}
\end{table}

\begin{table}[h]
\begin{center}
\begin{tabular}{c|c} 
SSB pattern & decomposition \\ [1ex] 
\hline\hline
$SU(3) \rightarrow A_4$ & ${\bf 15 \rightarrow 1 + (1'+1''+3+3)_{_G}+(3+3)_{_{deg.}}+1_{_{pGB}} + (1'+1'')_{_{deg.}}+3+3+3+3}$\\[2ex]
$SU(3) \rightarrow T_7$& ${\bf 15 \longrightarrow 1 + (1'+1''+3'+3'')_{_G}+(3'+3'')_{_{deg.}}}$\\[2ex]
&${\bf +1_{_{pGB}} + (1'+1'')_{_{deg.}}+(3'+3'')_{_{deg.}}+(3'+3'')_{_{deg.}}}$\\[2ex]
$SU(3) \rightarrow \Delta(27)$& ${\bf 10 \rightarrow 1 +1_{_{pGB}}  +(\Sigma_{n=2}^9)_{_G}+ 3+(1+1)_{_{pGB}} +(1_2+1_3)_{_{deg.}} +(\Sigma_{n=4}^9)_{_{deg.}}}$\\[2ex]
$SU(3) \rightarrow PSL(2,7)$& ${\bf 15' \rightarrow 1 + 6+8_{_G}+1_{_{pGB}} + 6 + 8}$\\[2ex]
$SU(3) \rightarrow PSL(2,7)$& ${\bf 28 \rightarrow 1 +6+6+7+ 8_{_G}+  1_{_{pGB}} + 6+6+7_{_{pGB}}+ 8}$\\[2ex]
\end{tabular}
\caption{Scalar mass eigenstates for the SSB patterns $SU(3) \rightarrow A_4,$ $T_7$,  $\Delta(27)$ and $PSL(2,7)$ using various complex  irreps of $SU(3)$.}
\label{tab:SU(3)sum}
\end{center}
\end{table}

\newpage
\section{Discussion and Conclusion}

The standard model includes  28 unspecified parameters, some of which describe fermion masses and mixing angles. Consequently, we do not know why the quark and lepton masses and mixings are what they are. To fix these parameters, a standard approach has been to extend the SM by a discrete symmetry, but this approach is not without its difficulties as discussed above. What would seem more natural would be to increase the gauge group to $SU(3)\times SU(2)\times U(1)\times G$ and extend the scalar sector. Then this model can be of the same general type as the SM, i. e., an anomaly-free gauge theory with fermions that gets spontaneously broken by VEVs of scalar fields. If the SSB of $G$ results in a discrete subgroup $\Gamma$ then we arrive at a $SU(3)\times SU(2)\times U(1)\times \Gamma$ via a route that avoids the problems just mentioned, without choosing an {\it ad hoc} discrete group for extending the SM. 

Here, based on the techniques of Luhn\cite{Luhn:2011ip} and Merle and R.~Zwicky\cite{Merle:2011vy}, we have demonstrated that we can carry out the $G\rightarrow \Gamma$ SSB in many cases of interest, specifically breaking to $A_4,S_4, A_5,Q_6,T',O',I',T_7,\Delta(27)$ and $PSL(2,7)$. Other cases can be handled by the same techniques. Many other discrete groups have been occasionally used to extend the SM, e.g., $D_4, D_5, D_7, D_{14}, \Delta(54), \Delta(96),$ and  $\Sigma(81)$ have all appeared in the literature \cite{Hagedorn:2008bc,King:2012in,Vien:2014ica}. For a discussion of breaking $SO(3)$ to dihedral groups see \cite{Koca:2003jy}. Further information about the classification of the discrete subgroups of $SU(3)$ can be found in \cite{ Ludl:2010bj,Grimus:2011fk,Ludl:2011gn,Merle:2011vy}.
In addition products of discrete groups are often employed, where the products often contain $Z_n$ factors. To gauge these cases we can start with a product gauge group and break to the 
desired discrete group, $G_1\times G_2\times ... \rightarrow \Gamma_1\times \Gamma_2\times ...$. As long as there are no cross terms in the scalar potential, then we can proceed as above. In some cases the cross terms can destabilize  the minima, so they must either be eliminated,  or dealt with by other means. If the fundamental charge of a $U(1)$ gauge group is $q$, then by breaking a $U(1)$ with scalar particle of charge $nq$ one arrives at $Z_n$.   Results given here could be applied to extend recent work on gauging two Higgs doublet models \cite{Huang:2015wts}.
 Using our results to extend models currently in the literature can solve some existing problems, and the inclusion of new scalars in the spectrum may be of interest since some may be detectable either directly or indirectly depending on the details of the model. Such phenomenological investigations need to proceed on a model by model basis, and we plan to look at some specific examples in future work.

\section{Acknowledgements} 

We have benefited greatly from discussions and correspondences with Christoph Luhn, Alex Merle and Pierre Ramond.

\newpage
\appendix
\section{Branching Rules}\label{app}

In this Appendix we present the branching rules for the embeddings of discrete groups into Lie groups used in the paper. The vertical axes label the dimensions of the Lie Group reps, and the horizontal the dimensions of the discrete group representations.

\begin{table}[h]
\parbox{.45\linewidth}{
\caption{$SO(3)\rightarrow A_4$}
\label{tab:4}
\centering
\begin{tabular}{c|cccc}
Dimension & $\mathbf{1_1}$&$\mathbf{1_2}$&$\mathbf{1_3}$&$\mathbf{3}$\\
\hline
\textbf{2}&0&0&0&0\\
\textbf{3}&0&0&0&1\\
\textbf{4}&0&0&0&0\\
\textbf{5}&0&1&1&1\\
\textbf{6}&0&0&0&0\\
\textbf{7}&1&0&0&2\\
\textbf{8}&0&0&0&0\\
\textbf{9}&1&1&1&2\\
\textbf{10}&0&0&0&0\\
\textbf{11}&0&1&1&3
\end{tabular}
}
 \hfill
\parbox{.45\linewidth}{
\caption{$SO(3)\rightarrow S_4$}
\label{tab:10}
\centering
\begin{tabular}{c|ccccc}
Dimension & $\mathbf{1_1}$&$\mathbf{1_2}$&$\mathbf{2}$&$\mathbf{3_1}$&$\mathbf{3_2}$\\
\hline
\textbf{2}&0&0&0&0&0\\
\textbf{3}&0&0&0&1&0\\
\textbf{4}&0&0&0&0&0\\
\textbf{5}&0&0&1&0&1\\
\textbf{6}&0&0&0&0&0\\
\textbf{7}&0&1&0&1&1\\
\textbf{8}&0&0&0&0&0\\
\textbf{9}&1&0&1&1&1\\
\textbf{10}&0&0&0&0&0\\
\textbf{11}&0&0&1&2&1
\end{tabular}
}
\end{table}

\begin{table}[h]
\parbox{.45\linewidth}{
\caption{$SO(3)\rightarrow A_5$}
\label{tab:3}
\centering
\begin{tabular}{c|ccccc}
Dimension & $\mathbf{1}$&$\mathbf{3}$&$\mathbf{3}$&$\mathbf{4}$&$\mathbf{5}$\\
\hline
\textbf{2}&0&0&0&0&0\\
\textbf{3}&0&0&1&0&0\\
\textbf{4}&0&0&0&0&0\\
\textbf{5}&0&0&0&0&1\\
\textbf{6}&0&0&0&0&0\\
\textbf{7}&0&1&0&1&0\\
\textbf{8}&0&0&0&0&0\\
\textbf{9}&0&0&0&1&1\\
*\textbf{10}&0&0&0&0&0\\
\textbf{11}&0&1&1&0&1\\
\textbf{12}&0&0&0&0&0\\
\textbf{13}&1&0&1&1&1
\end{tabular}
}
\hfill
\parbox{.45\linewidth}{
\caption{$SU(2)\rightarrow Q_6$}
\label{tab:1}
\centering
\begin{tabular}{c|cccccc}
Dimension & $\mathbf{1_1}$&$\mathbf{1_2}$&$\mathbf{1_3}$&$\mathbf{1_4}$&$\mathbf{2_1}$&$\mathbf{2_2}$\\
\hline
\textbf{2}&0&0&0&0&1&0\\
\textbf{3}&0&1&0&0&0&1\\
\textbf{4}&0&0&1&1&1&0\\
\textbf{5}&1&0&0&0&0&2\\
\textbf{6}&0&0&1&1&2&0\\
\textbf{7}&1&2&0&0&0&2\\
\textbf{8}&0&0&1&1&3&0\\
\textbf{9}&2&1&0&0&0&3\\
\textbf{10}&0&0&2&2&3&0\\
\textbf{11}&1&2&0&0&0&4
\end{tabular}
}
\end{table}

\begin{table}[h]
\parbox{.45\linewidth}{
\caption{$SU(2)\rightarrow T'$}
\label{tab:2}
\centering
\begin{tabular}{c|ccccccc}
Dimension & $\mathbf{1_1}$&$\mathbf{1_2}$&$\mathbf{1_3}$&$\mathbf{2_1}$&$\mathbf{2_2}$&$\mathbf{2_3}$&$\mathbf{3}$\\
\hline
\textbf{2}&0&0&0&1&0&0&0\\
\textbf{3}&0&0&0&0&0&0&1\\
\textbf{4}&0&0&0&0&1&1&0\\
\textbf{5}&0&1&1&0&0&0&1\\
\textbf{6}&0&0&0&1&1&1&0\\
\textbf{7}&1&0&0&0&0&0&2\\
\textbf{8}&0&0&0&2&1&1&0\\
\textbf{9}&1&1&1&0&0&0&2\\
\textbf{10}&0&0&0&1&2&2&0\\
\textbf{11}&0&1&1&0&0&0&3
\end{tabular}
}
\hfill
\parbox{.45\linewidth}{
\caption{$SU(2)\rightarrow O'$}
\label{tab:11}
\centering
\begin{tabular}{c|cccccccc}
Dimension & $\mathbf{1_1}$&$\mathbf{1_2}$&$\mathbf{2_1}$&$\mathbf{2_2}$&$\mathbf{2_3}$&$\mathbf{3_1}$&$\mathbf{3_2}$&$\mathbf{4}$\\
\hline
\textbf{2}&0&0&0&1&0&0&0&0\\
\textbf{3}&0&0&0&0&0&0&1&0\\
\textbf{4}&0&0&0&0&0&0&0&1\\
\textbf{5}&0&0&1&0&0&1&0&0\\
\textbf{6}&0&0&0&0&1&0&0&1\\
\textbf{7}&0&1&0&0&0&1&1&0\\
\textbf{8}&0&0&0&1&1&0&0&1\\
\textbf{9}&1&0&1&0&0&1&1&0\\
\textbf{10}&0&0&0&1&0&0&0&2\\
\textbf{11}&0&0&1&0&0&1&2&0
\end{tabular}

}
\end{table}

\begin{table}[h]
\parbox{.45\linewidth}{
\caption{$SU(2)\rightarrow I'$}
\label{tab:5}
\centering
\begin{tabular}{c|ccccccccc}
Dimension & $\mathbf{1_1}$&$\mathbf{2_2}$&$\mathbf{2_3}$&$\mathbf{3_1}$&$\mathbf{3_2}$&$\mathbf{4_1}$&$\mathbf{4_2}$&$\mathbf{5}$&$\mathbf{6}$\\
\hline
\textbf{2}&0&1&0&0&0&0&0&0&0\\
\textbf{3}&0&0&0&0&1&0&0&0&0\\
\textbf{4}&0&0&0&0&0&0&1&0&0\\
\textbf{5}&0&0&0&0&0&0&0&1&0\\
\textbf{6}&0&0&0&0&0&0&0&0&1\\
\textbf{7}&0&0&0&1&0&1&0&0&0\\
\textbf{8}&0&0&1&0&0&0&0&0&1\\
\textbf{9}&0&0&0&0&0&1&0&1&0\\
\textbf{10}&0&1&0&0&0&0&1&0&1\\
\textbf{11}&0&0&0&1&1&0&0&1&0\\
\textbf{12}&0&1&0&0&0&0&1&0&1\\
\textbf{13}&1&0&0&0&1&1&0&1&0
\end{tabular}
}
\hfill
\parbox{.45\linewidth}{
\caption{$SU(3)\rightarrow A_4$}
\label{tab:6}
\centering
\begin{tabular}{c|cccc}
Dimension & $\mathbf{1_1}$&$\mathbf{1_2}$&$\mathbf{1_3}$&$\mathbf{3}$\\
\hline
\textbf{3}&0&0&0&1\\
\textbf{6}&1&1&1&1\\
\textbf{8}&0&1&1&2\\
\textbf{10}&1&0&0&3\\
\textbf{15}&1&1&1&4\\
\textbf{15$\bm{'}$}&2&2&2&3\\
\textbf{21}&1&1&1&6\\
\textbf{24}&2&2&2&6\\
\textbf{27}&3&3&3&6
\end{tabular}
}
\end{table}

\begin{table}[h]
\parbox{.45\linewidth}{
\caption{$SU(3)\rightarrow T_7$ }
\label{tab:9}
\centering
\begin{tabular}{c|ccccc}
Dimension & $\mathbf{1_1}$&$\mathbf{1_2}$&$\mathbf{1_3}$&$\mathbf{3_1}$&$\mathbf{3_2}$\\
\hline
\textbf{3}&0&0&0&1&0\\
\textbf{6}&0&0&0&1&1\\
\textbf{8}&0&1&1&1&1\\
\textbf{10}&1&0&0&1&2\\
\textbf{15}&1&1&1&2&2\\
\textbf{15$\bm{'}$}&1&1&1&2&2\\
\textbf{21}&1&1&1&3&3\\
\textbf{24}&1&1&1&4&3\\
\textbf{27}&1&1&1&4&4
\end{tabular}
}
\hfill
\parbox{.45\linewidth}{
\caption{$SU(3)\rightarrow \Delta(27)$}
\label{tab:7}
\centering
\begin{tabular}{c|ccccccccccc}
Dimension & $\mathbf{1_1}$&$\mathbf{1_2}$&$\mathbf{1_3}$&$\mathbf{1_4}$&$\mathbf{1_5}$&$\mathbf{1_6}$&$\mathbf{1_7}$&$\mathbf{1_8}$&$\mathbf{1_9}$&$\mathbf{3_1}$&$\mathbf{3_2}$\\
\hline
\textbf{3}&0&0&0&0&0&0&0&0&0&1&0\\
\textbf{6}&0&0&0&0&0&0&0&0&0&0&2\\
\textbf{8}&0&1&1&1&1&1&1&1&1&0&0\\
\textbf{10}&2&1&1&1&1&1&1&1&1&0&0\\
\textbf{15}&0&0&0&0&0&0&0&0&0&5&0\\
\textbf{15$\bm{'}$}&0&0&0&0&0&0&0&0&0&5&0\\
\textbf{21}&0&0&0&0&0&0&0&0&0&0&7\\
\textbf{24}&0&0&0&0&0&0&0&0&0&0&8\\
\textbf{27}&3&3&3&3&3&3&3&3&3&0&0
\end{tabular}
}
\end{table}

\begin{table}[h]
\caption{$SU(3)\rightarrow PSL(2,7) $}
\label{tab:8}
\begin{center}
\begin{tabular}{c|cccccc}
Dimension & $\mathbf{1}$&$\mathbf{3_2}$&$\mathbf{3_2}$&$\mathbf{6}$&$\mathbf{7}$&$\mathbf{8}$\\
\hline
\textbf{3}&0&1&0&0&0&0\\
\textbf{6}&0&0&0&1&0&0\\
\textbf{8}&0&0&0&0&0&1\\
\textbf{10}&0&0&1&0&1&0\\
\textbf{15}&0&0&0&0&1&1\\
\textbf{15$\bm{'}$}&1&0&0&1&0&1\\
\textbf{21}&0&1&1&0&1&1\\
\textbf{24}&0&1&0&1&1&1\\
\textbf{27}&0&0&0&2&1&1\\
\textbf{28}&1&0&0&2&1&1
\end{tabular}
\end{center}
\end{table}


\begin{thebibliography}{99}

  %HISTORY

%\cite{Pakvasa:1977in}
\bibitem{Pakvasa:1977in} 
  S.~Pakvasa and H.~Sugawara,
  %``Discrete Symmetry and Cabibbo Angle,''
  Phys.\ Lett.\  {\bf 73B}, 61 (1978).
  doi:10.1016/0370-2693(78)90172-7
  %%CITATION = doi:10.1016/0370-2693(78)90172-7;%%
  %321 citations counted in INSPIRE as of 16 Feb 2017
  
  %\cite{Ma:2001dn}
\bibitem{Ma:2001dn} 
  E.~Ma and G.~Rajasekaran,
  %``Softly broken A(4) symmetry for nearly degenerate neutrino masses,''
  Phys.\ Rev.\ D {\bf 64}, 113012 (2001)
  doi:10.1103/PhysRevD.64.113012
  [hep-ph/0106291].
  %%CITATION = doi:10.1103/PhysRevD.64.113012;%%
  %605 citations counted in INSPIRE as of 16 Feb 2017
  
  %\cite{Babu:2002dz}
\bibitem{Babu:2002dz} 
  K.~S.~Babu, E.~Ma and J.~W.~F.~Valle,
  %``Underlying A(4) symmetry for the neutrino mass matrix and the quark mixing matrix,''
  Phys.\ Lett.\ B {\bf 552}, 207 (2003)
  doi:10.1016/S0370-2693(02)03153-2
  [hep-ph/0206292].
  %%CITATION = doi:10.1016/S0370-2693(02)03153-2;%%
  %590 citations counted in INSPIRE as of 16 Feb 2017
  


  %REVIEWS

  
\bibitem{Frampton:1994rk}
  P.~H.~Frampton and T.~W.~Kephart,
  %``Simple nonAbelian finite flavor groups and fermion masses,''
  Int.\ J.\ Mod.\ Phys.\  A {\bf 10}, 4689 (1995)
  [arXiv:hep-ph/9409330].
  
\bibitem{Ishimori:2010au}
  H.~Ishimori, T.~Kobayashi, H.~Ohki, H.~Okada, Y.~Shimizu and M.~Tanimoto,
  %``Non-Abelian Discrete Symmetries in Particle Physics,''
  Prog.\ Theor.\ Phys.\ Suppl.\  {\bf 183}, 1 (2010)
  [arXiv:1003.3552 [hep-th]].  
  
  %\cite{Altarelli:2010gt}
\bibitem{Altarelli:2010gt} 
  G.~Altarelli and F.~Feruglio,
  %``Discrete Flavor Symmetries and Models of Neutrino Mixing,''
  Rev.\ Mod.\ Phys.\  {\bf 82}, 2701 (2010)
  doi:10.1103/RevModPhys.82.2701
  [arXiv:1002.0211 [hep-ph]].
  %%CITATION = doi:10.1103/RevModPhys.82.2701;%%
  %495 citations counted in INSPIRE as of 09 Oct 2016
  
    
 %\cite{King:2013eh}
\bibitem{King:2013eh} 
  S.~F.~King and C.~Luhn,
  %``Neutrino Mass and Mixing with Discrete Symmetry,''
  Rept.\ Prog.\ Phys.\  {\bf 76}, 056201 (2013)
  doi:10.1088/0034-4885/76/5/056201
  [arXiv:1301.1340 [hep-ph]].
  %%CITATION = doi:10.1088/0034-4885/76/5/056201;%%
  %331 citations counted in INSPIRE as of 08 Feb 2017
  
  %\cite{King:2014nza}
\bibitem{King:2014nza} 
  S.~F.~King, A.~Merle, S.~Morisi, Y.~Shimizu and M.~Tanimoto,
  %``Neutrino Mass and Mixing: from Theory to Experiment,''
  New J.\ Phys.\  {\bf 16}, 045018 (2014)
  doi:10.1088/1367-2630/16/4/045018
  [arXiv:1402.4271 [hep-ph]].
  %%CITATION = doi:10.1088/1367-2630/16/4/045018;%%
  %150 citations counted in INSPIRE as of 23 Feb 2017
  
  %\cite{Krauss:1988zc}
\bibitem{Krauss:1988zc} 
  L.~M.~Krauss and F.~Wilczek,
  %``Discrete Gauge Symmetry in Continuum Theories,''
  Phys.\ Rev.\ Lett.\  {\bf 62}, 1221 (1989).
  doi:10.1103/PhysRevLett.62.1221
  %%CITATION = doi:10.1103/PhysRevLett.62.1221;%%
  %488 citations counted in INSPIRE as of 16 Feb 2017
  
  %\cite{Luhn:2008sa}
\bibitem{Luhn:2008sa} 
  C.~Luhn and P.~Ramond,
  %``Anomaly Conditions for Non-Abelian Finite Family Symmetries,''
  JHEP {\bf 0807}, 085 (2008)
  doi:10.1088/1126-6708/2008/07/085
  [arXiv:0805.1736 [hep-ph]].
  %%CITATION = doi:10.1088/1126-6708/2008/07/085;%%
  %32 citations counted in INSPIRE as of 16 Feb 2017
  
  %\cite{Hindmarsh:1994re}
\bibitem{Hindmarsh:1994re} 
  M.~B.~Hindmarsh and T.~W.~B.~Kibble,
  %``Cosmic strings,''
  Rept.\ Prog.\ Phys.\  {\bf 58}, 477 (1995)
  doi:10.1088/0034-4885/58/5/001
  [hep-ph/9411342].
  %%CITATION = doi:10.1088/0034-4885/58/5/001;%%
  %735 citations counted in INSPIRE as of 16 Feb 2017
  
  
    %\cite{Reynolds}
\bibitem{Reynolds} 
  O. Reynolds,
  %"On the dynamical theory of incompressible viscous fluids and the determination of the criterion", 
  Phil. Trans. of the Royal Society A, 186: 123 (1895);
  B. Sturmfels, {\it Algorithms in Invariant Theory,} $2^{nd}$ ed., (2008) Springer, Wien \& New York.
  
 %\cite{Molien}
  \bibitem{Molien} 
  T. Molien, 
  %Uber die Invarianten der linearen Substitutionsgruppen, 
  Sitz. K\"onig Preuss. Akad.
Wiss. (1897), N 52, 1152-1156.

  %\cite{Luhn:2011ip}
\bibitem{Luhn:2011ip} 
  C.~Luhn,
  %``Spontaneous breaking of SU(3) to finite family symmetries: a pedestrian's approach,''
  JHEP {\bf 1103}, 108 (2011)
  doi:10.1007/JHEP03(2011)108
  [arXiv:1101.2417 [hep-ph]].
  %%CITATION = doi:10.1007/JHEP03(2011)108;%%
  %25 citations counted in INSPIRE as of 19 Oct 2016
  
  %\cite{Merle:2011vy}
\bibitem{Merle:2011vy} 
  A.~Merle and R.~Zwicky,
  %``Explicit and spontaneous breaking of SU(3) into its finite subgroups,''
  JHEP {\bf 1202}, 128 (2012)
  doi:10.1007/JHEP02(2012)128
  [arXiv:1110.4891 [hep-ph]].
  %%CITATION = doi:10.1007/JHEP02(2012)128;%%
  %27 citations counted in INSPIRE as of 06 Oct 2016
  


  %\cite{Fallbacher:2015pga
\bibitem{Fallbacher:2015pga} 
  M.~Fallbacher,
  %``Breaking classical Lie groups to finite subgroups Ð an automated approach,''
  Nucl.\ Phys.\ B {\bf 898}, 229 (2015)
  doi:10.1016/j.nuclphysb.2015.07.004
  [arXiv:1506.03677 [hep-th]].
  %%CITATION = doi:10.1016/j.nuclphysb.2015.07.004;%%
  %2 citations counted in INSPIRE as of 06 Oct 2016
  
 \bibitem{GAP4}  GAP - ``Groups, Algorithms, Programming -
a System for Computational Discrete Algebra,''
version GAP 4.8.5, 25 Sept. 2016. 

 
  
  %REP THEORY
    \bibitem{Frampton:2009pr}
  P.~H.~Frampton, T.~W.~Kephart and R.~M.~Rohm,
  %``A Note on Embedding Nonabelian Finite Flavor Groups in Continuous Groups,''
  Phys.\ Lett.\  B {\bf 679}, 478 (2009)
  [arXiv:0904.0420 [hep-ph]].
  %%CITATION = PHLTA,B679,478;%%
  
    %A4
  
    %\cite{Berger:2009tt}
\bibitem{Berger:2009tt} 
  J.~Berger and Y.~Grossman,
  %``Model of leptons from SO(3) ---> A(4),''
  JHEP {\bf 1002}, 071 (2010)
  doi:10.1007/JHEP02(2010)071
  [arXiv:0910.4392 [hep-ph]].
  %%CITATION = doi:10.1007/JHEP02(2010)071;%%
  %43 citations counted in INSPIRE as of 09 Oct 2016

%\cite{Ma:2015fpa}
\bibitem{Ma:2015fpa} 
  E.~Ma,
  %``Neutrino Mixing: $A_4$ Variations,''
  Phys.\ Lett.\ B {\bf 752}, 198 (2016)
  doi:10.1016/j.physletb.2015.11.049
  [arXiv:1510.02501 [hep-ph]].
  %%CITATION = doi:10.1016/j.physletb.2015.11.049;%%
  %10 citations counted in INSPIRE as of 19 Feb 2017



  
    %\cite{Altarelli:2008bg}
\bibitem{Altarelli:2008bg}
  G.~Altarelli, F.~Feruglio and C.~Hagedorn,
  %``A SUSY SU(5) Grand Unified Model of Tri-Bimaximal Mixing from A$_4$,''
  JHEP {\bf 0803} (2008) 052
  doi:10.1088/1126-6708/2008/03/052
  [arXiv:0802.0090 [hep-ph]].
  %%CITATION = doi:10.1088/1126-6708/2008/03/052;%%
  %162 citations counted in INSPIRE as of 18 Feb 2017
  

  
  %\cite{King:2011ab}
\bibitem{King:2011ab} 
  S.~F.~King and C.~Luhn,
  %``A4 models of tri-bimaximal-reactor mixing,''
  JHEP {\bf 1203}, 036 (2012)
  doi:10.1007/JHEP03(2012)036
  [arXiv:1112.1959 [hep-ph]].
  %%CITATION = doi:10.1007/JHEP03(2012)036;%%
  %52 citations counted in INSPIRE as of 08 Feb 2017
  
   %\cite{Ferreira:2013oga}
\bibitem{Ferreira:2013oga} 
  P.~M.~Ferreira, L.~Lavoura and P.~O.~Ludl,
  %``A new $A_{4}$ model for lepton mixing,''
  Phys.\ Lett.\ B {\bf 726}, 767 (2013)
  doi:10.1016/j.physletb.2013.09.058
  [arXiv:1306.1500 [hep-ph]].
  %%CITATION = doi:10.1016/j.physletb.2013.09.058;%%
  %17 citations counted in INSPIRE as of 08 Feb 2017

  
%\cite{Cvitanovic:1976am}
\bibitem{Cvitanovic:1976am} 
  P.~Cvitanovic,
  %``Group theory for Feynman diagrams in non-Abelian gauge theories,''
  Phys.\ Rev.\ D {\bf 14}, 1536 (1976).
  doi:10.1103/PhysRevD.14.1536
  %%CITATION = doi:10.1103/PhysRevD.14.1536;%%
  %191 citations counted in INSPIRE as of 09 Oct 2016
  
 %S4
 
 %\cite{Hagedorn:2006ug}
\bibitem{Hagedorn:2006ug} 
  C.~Hagedorn, M.~Lindner and R.~N.~Mohapatra,
  %``S(4) flavor symmetry and fermion masses: Towards a grand unified theory of flavor,''
  JHEP {\bf 0606}, 042 (2006)
  doi:10.1088/1126-6708/2006/06/042
  [hep-ph/0602244].
  %%CITATION = doi:10.1088/1126-6708/2006/06/042;%%
  %182 citations counted in INSPIRE as of 18 Feb 2017
  
 %\cite{Hagedorn:2010th}
\bibitem{Hagedorn:2010th} 
  C.~Hagedorn, S.~F.~King and C.~Luhn,
  %``A SUSY GUT of Flavour with S4 x SU(5) to NLO,''
  JHEP {\bf 1006}, 048 (2010)
  doi:10.1007/JHEP06(2010)048
  [arXiv:1003.4249 [hep-ph]].
  %%CITATION = doi:10.1007/JHEP06(2010)048;%%
  %93 citations counted in INSPIRE as of 18 Feb 2017
  %A5

%\cite{Everett:2008et}
\bibitem{Everett:2008et} 
  L.~L.~Everett and A.~J.~Stuart,
  %``Icosahedral (A(5)) Family Symmetry and the Golden Ratio Prediction for Solar Neutrino Mixing,''
  Phys.\ Rev.\ D {\bf 79}, 085005 (2009)
  [arXiv:0812.1057 [hep-ph]];
  
  %\cite{Feruglio:2011qq}
\bibitem{Feruglio:2011qq} 
  F.~Feruglio and A.~Paris,
  %``The Golden Ratio Prediction for the Solar Angle from a Natural Model with A5 Flavour Symmetry,''
  JHEP {\bf 1103}, 101 (2011)
  [arXiv:1101.0393 [hep-ph]];
  %%CITATION = ARXIV:1101.0393;%%
  %31 citations counted in INSPIRE as of 05 Mar 2013
  %%CITATION = ARXIV:0812.1057;%%
  %56 citations counted in INSPIRE as of 05 Mar 2013
  
%\cite{Ding:2011cm}
\bibitem{Ding:2011cm} 
  G.~-J.~Ding, L.~L.~Everett and A.~J.~Stuart,
  %``Golden Ratio Neutrino Mixing and $A_5$ Flavor Symmetry,''
  Nucl.\ Phys.\ B {\bf 857}, 219 (2012)
  [arXiv:1110.1688 [hep-ph]];
  
  \bibitem{Chen:2010ty}
  C.~-S.~Chen, T.~W.~Kephart and T.~-C.~Yuan,
  %``An A_5 Model of Four Lepton Generations,''
  JHEP {\bf 1104}, 015 (2011)
  [arXiv:1011.3199 [hep-ph]].
  
  %Q6
  
  %\cite{Frampton:1994xm}
\bibitem{Frampton:1994xm} 
  P.~H.~Frampton and T.~W.~Kephart,
  %``Minimal family unification,''
  Phys.\ Rev.\ D {\bf 51}, R1 (1995)
  doi:10.1103/PhysRevD.51.R1
  [hep-ph/9409324].
  %%CITATION = doi:10.1103/PhysRevD.51.R1;%%
  %41 citations counted in INSPIRE as of 08 Feb 2017
  
%\cite{Frampton:1999hk}
\bibitem{Frampton:1999hk} 
  P.~H.~Frampton and A.~Rasin,
  %``NonAbelian discrete symmetries, fermion mass textures and large neutrino mixing,''
  Phys.\ Lett.\ B {\bf 478}, 424 (2000)
  doi:10.1016/S0370-2693(00)00276-8
  [hep-ph/9910522].
  %%CITATION = doi:10.1016/S0370-2693(00)00276-8;%%
  %67 citations counted in INSPIRE as of 16 Dec 2016
  
  %T '

  
  %\cite{Aranda:1999kc}
\bibitem{Aranda:1999kc} 
  A.~Aranda, C.~D.~Carone and R.~F.~Lebed,
  %``U(2) flavor physics without U(2) symmetry,''
  Phys.\ Lett.\ B {\bf 474}, 170 (2000)
  doi:10.1016/S0370-2693(99)01497-5
  [hep-ph/9910392].
  %%CITATION = doi:10.1016/S0370-2693(99)01497-5;%%
  %80 citations counted in INSPIRE as of 16 Feb 2017
  
   %\cite{Chen:2007afa}
\bibitem{Chen:2007afa} 
  M.~C.~Chen and K.~T.~Mahanthappa,
  %``CKM and Tri-bimaximal MNS Matrices in a $SU(5) \times ^{(d)}T$ Model,''
  Phys.\ Lett.\ B {\bf 652}, 34 (2007)
  doi:10.1016/j.physletb.2007.06.064
  [arXiv:0705.0714 [hep-ph]].
  %%CITATION = doi:10.1016/j.physletb.2007.06.064;%%
  %207 citations counted in INSPIRE as of 19 Feb 2017
  
  \bibitem{Frampton:2007et}
  P.~H.~Frampton and T.~W.~Kephart, %``Flavor Symmetry for Quarks and Leptons,''
  JHEP {\bf 0709}, 110 (2007)
  [arXiv:0706.1186 [hep-ph]];
 % \bibitem{Frampton:2008bz}
 

 
%\cite{Frampton:2008bz}
\bibitem{Frampton:2008bz} 
  P.~H.~Frampton, T.~W.~Kephart and S.~Matsuzaki,
  %``Simplified Renormalizable T-prime Model for Tribimaximal Mixing and Cabibbo Angle,''
  Phys.\ Rev.\ D {\bf 78}, 073004 (2008)
  doi:10.1103/PhysRevD.78.073004
  [arXiv:0807.4713 [hep-ph]].
  %%CITATION = doi:10.1103/PhysRevD.78.073004;%%
  %55 citations counted in INSPIRE as of 18 Feb 2017
  
  %\cite{Frampton:2010uw}
\bibitem{Frampton:2010uw} 
  P.~H.~Frampton, C.~M.~Ho, T.~W.~Kephart and S.~Matsuzaki,
  %``LHC Higgs Production and Decay in the $T'$ Model,''
  Phys.\ Rev.\ D {\bf 82}, 113007 (2010)
  doi:10.1103/PhysRevD.82.113007
  [arXiv:1009.0307 [hep-ph]].
  %%CITATION = doi:10.1103/PhysRevD.82.113007;%%
  %14 citations counted in INSPIRE as of 18 Feb 2017

  %\cite{Natale:2016xob}
\bibitem{Natale:2016xob} 
  A.~Natale,
  %``A Radiative Model of Quark Masses with Binary Tetrahedral Symmetry,''
  Nucl.\ Phys.\ B {\bf 914}, 201 (2017)
  doi:10.1016/j.nuclphysb.2016.11.006
  [arXiv:1608.06999 [hep-ph]].
  %%CITATION = doi:10.1016/j.nuclphysb.2016.11.006;%%
  
   %\cite{Carone:2016xsi}
\bibitem{Carone:2016xsi} 
  C.~D.~Carone, S.~Chaurasia and S.~Vasquez,
  %``Flavor from the double tetrahedral group without supersymmetry,''
  Phys.\ Rev.\ D {\bf 95}, no. 1, 015025 (2017)
  doi:10.1103/PhysRevD.95.015025
  [arXiv:1611.00784 [hep-ph]].
  %%CITATION = doi:10.1103/PhysRevD.95.015025;%%
  
 % O'
  
  %I '
  
 
  
  %\cite{Everett:2010rd}
\bibitem{Everett:2010rd} 
  L.~L.~Everett and A.~J.~Stuart,
  %``The Double Cover of the Icosahedral Symmetry Group and Quark Mass Textures,''
  Phys.\ Lett.\ B {\bf 698}, 131 (2011)
  doi:10.1016/j.physletb.2011.02.054
  [arXiv:1011.4928 [hep-ph]].
  %%CITATION = doi:10.1016/j.physletb.2011.02.054;%%
  %20 citations counted in INSPIRE as of 09 Oct 2016
  
  %\cite{Chen:2011dn}
\bibitem{Chen:2011dn} 
  C.~S.~Chen, T.~W.~Kephart and T.~C.~Yuan,
  %``Binary Icosahedral Flavor Symmetry for Four Generations of Quarks and Leptons,''
  PTEP {\bf 2013}, no. 10, 103B01 (2013)
  doi:10.1093/ptep/ptt071
  [arXiv:1110.6233 [hep-ph]].
  %%CITATION = doi:10.1093/ptep/ptt071;%%
  %6 citations counted in INSPIRE as of 09 Oct 2016
  

  
  %T7
  %\cite{Luhn:2007sy}
\bibitem{Luhn:2007sy} 
  C.~Luhn, S.~Nasri and P.~Ramond,
  %``Tri-bimaximal neutrino mixing and the family symmetry semidirect product of Z(7) and Z(3),''
  Phys.\ Lett.\ B {\bf 652}, 27 (2007)
  doi:10.1016/j.physletb.2007.06.059
  [arXiv:0706.2341 [hep-ph]].
  %%CITATION = doi:10.1016/j.physletb.2007.06.059;%%
  %145 citations counted in INSPIRE as of 09 Oct 2016
  
  %\cite{Kile:2014kya}
\bibitem{Kile:2014kya} 
  J.~Kile, M.~J.~PŽrez, P.~Ramond and J.~Zhang,
  %``$?_{13}$ and the flavor ring,''
  Phys.\ Rev.\ D {\bf 90}, no. 1, 013004 (2014)
  doi:10.1103/PhysRevD.90.013004
  [arXiv:1403.6136 [hep-ph]].
  %%CITATION = doi:10.1103/PhysRevD.90.013004;%%
  %7 citations counted in INSPIRE as of 09 Oct 2016
  

  
    %\cite{Vien:2015koa}
\bibitem{Vien:2015koa} 
  V.~V.~Vien,
  %``$T_7$ flavor symmetry scheme for understanding neutrino mass and mixing in 3-3-1 model with neutral leptons,''
  Mod.\ Phys.\ Lett.\ A {\bf 29}, 28 (2014)
  doi:10.1142/S0217732314501399
  [arXiv:1508.02585 [hep-ph]].
  %%CITATION = doi:10.1142/S0217732314501399;%%
  %3 citations counted in INSPIRE as of 08 Feb 2017
  
  %\cite{Vien:2016qbb}
\bibitem{Vien:2016qbb} 
  V.~V.~Vien and H.~N.~Long,
  %``Lepton mass and mixing in a simple extension of the Standard Model based on T7 flavor symmetry,''
  arXiv:1609.03895 [hep-ph].
  %%CITATION = ARXIV:1609.03895;%%
  %1 citations counted in INSPIRE as of 08 Feb 2017
  
    %Delta(27)

  %\cite{Vien:2016tmh}
\bibitem{Vien:2016tmh} 
  V.~V.~Vien, A.~E.~C‡rcamo Hern‡ndez and H.~N.~Long,
  %``The $\Delta(27)$ flavor 3-3-1 model with neutral leptons,''
  Nucl.\ Phys.\ B {\bf 913}, 792 (2016)
  doi:10.1016/j.nuclphysb.2016.10.010
  [arXiv:1601.03300 [hep-ph]].
  %%CITATION = doi:10.1016/j.nuclphysb.2016.10.010;%%
  %14 citations counted in INSPIRE as of 08 Feb 2017
  
  %\cite{Ferreira:2012ri}
\bibitem{Ferreira:2012ri} 
  P.~M.~Ferreira, W.~Grimus, L.~Lavoura and P.~O.~Ludl,
  %``Maximal CP Violation in Lepton Mixing from a Model with Delta(27) flavour Symmetry,''
  JHEP {\bf 1209}, 128 (2012)
  doi:10.1007/JHEP09(2012)128
  [arXiv:1206.7072 [hep-ph]].
  %%CITATION = doi:10.1007/JHEP09(2012)128;%%
  %48 citations counted in INSPIRE as of 08 Feb 2017
  
 
  %PSL(2,7)
  
  %\cite{Chen:2014wiw}
\bibitem{Chen:2014wiw} 
  G.~Chen, M.~J.~PŽrez and P.~Ramond,
  %``Neutrino masses, the $\mu$-term and $\mathcal{ PSL}_2(7)$,''
  Phys.\ Rev.\ D {\bf 92}, no. 7, 076006 (2015)
  doi:10.1103/PhysRevD.92.076006
  [arXiv:1412.6107 [hep-ph]].
  %%CITATION = doi:10.1103/PhysRevD.92.076006;%%
  %5 citations counted in INSPIRE as of 09 Oct 2016
  




 

    % Other discrete groups,  D4, D5, D7, D14, Delta54, Delta96,  Sigma(81)
 
   % Dn
%\cite{Koca:2003jy}
\bibitem{Koca:2003jy} 
  M.~Koca, R.~Koc and H.~Tutunculer,
  %``Explicit breaking of SO(3) with Higgs fields in the representations l = 2 and l = 3,''
  Int.\ J.\ Mod.\ Phys.\ A {\bf 18}, 4817 (2003)
  doi:10.1142/S0217751X03015891
  [hep-ph/0410270].
  %%CITATION = doi:10.1142/S0217751X03015891;%%
  %8 citations counted in INSPIRE as of 27 Oct 2016
  
  \bibitem{C.Luhn} Christoph Luhn, private communication.
  
  %\cite{Hagedorn:2008bc}
\bibitem{Hagedorn:2008bc} 
  C.~Hagedorn, M.~A.~Schmidt and A.~Y.~Smirnov,
  %``Lepton Mixing and Cancellation of the Dirac Mass Hierarchy in SO(10) GUTs with Flavor Symmetries T(7) and Sigma(81),''
  Phys.\ Rev.\ D {\bf 79}, 036002 (2009)
  doi:10.1103/PhysRevD.79.036002
  [arXiv:0811.2955 [hep-ph]].
  %%CITATION = doi:10.1103/PhysRevD.79.036002;%%
  %67 citations counted in INSPIRE as of 19 Feb 2017
  
  %\cite{King:2012in}
\bibitem{King:2012in} 
  S.~F.~King, C.~Luhn and A.~J.~Stuart,
  %``A Grand Delta(96) x SU(5) Flavour Model,''
  Nucl.\ Phys.\ B {\bf 867}, 203 (2013)
  doi:10.1016/j.nuclphysb.2012.09.021
  [arXiv:1207.5741 [hep-ph]].
  %%CITATION = doi:10.1016/j.nuclphysb.2012.09.021;%%
  %65 citations counted in INSPIRE as of 08 Feb 2017
  
%\cite{Vien:2014ica}
\bibitem{Vien:2014ica} 
  V.~V.~Vien and H.~N.~Long,
  %``Quark masses and mixings in the 3-3-1 model with neutral leptons based on $D_{4}$ flavor symmetry,''
  J.\ Korean Phys.\ Soc.\  {\bf 66}, no. 12, 1809 (2015)
  doi:10.3938/jkps.66.1809
  [arXiv:1408.4333 [hep-ph]].
  %%CITATION = doi:10.3938/jkps.66.1809;%%
  %7 citations counted in INSPIRE as of 08 Feb 2017
  

  
  %\cite{Ludl:2010bj}
\bibitem{Ludl:2010bj} 
  P.~O.~Ludl,
  %``On the finite subgroups of U(3) of order smaller than 512,''
  J.\ Phys.\ A {\bf 43}, 395204 (2010)
  Erratum: [J.\ Phys.\ A {\bf 44}, 139501 (2011)]
  doi:10.1088/1751-8113/44/13/139501, 10.1088/1751-8113/43/39/395204
  [arXiv:1006.1479 [math-ph]].
  %%CITATION = doi:10.1088/1751-8113/44/13/139501, 10.1088/1751-8113/43/39/395204;%%
  %32 citations counted in INSPIRE as of 08 Feb 2017
  
    %\cite{Grimus:2011fk}
\bibitem{Grimus:2011fk} 
  W.~Grimus and P.~O.~Ludl,
  %``Finite flavour groups of fermions,''
  J.\ Phys.\ A {\bf 45}, 233001 (2012)
  doi:10.1088/1751-8113/45/23/233001
  [arXiv:1110.6376 [hep-ph]].
  %%CITATION = doi:10.1088/1751-8113/45/23/233001;%%
  %93 citations counted in INSPIRE as of 08 Feb 2017
  
  %\cite{Ludl:2011gn}
\bibitem{Ludl:2011gn} 
  P.~O.~Ludl,
  %``Comments on the classification of the finite subgroups of SU(3),''
  J.\ Phys.\ A {\bf 44}, 255204 (2011)
  Erratum: [J.\ Phys.\ A {\bf 45}, 069502 (2012)]
  doi:10.1088/1751-8113/45/6/069502, 10.1088/1751-8113/44/25/255204
  [arXiv:1101.2308 [math-ph]].
  %%CITATION = doi:10.1088/1751-8113/45/6/069502, 10.1088/1751-8113/44/25/255204;%%
  %18 citations counted in INSPIRE as of 08 Feb 2017

  %\cite{Huang:2015wts}
\bibitem{Huang:2015wts} 
  W.~C.~Huang, Y.~L.~S.~Tsai and T.~C.~Yuan,
  %``G2HDM : Gauged Two Higgs Doublet Model,''
  JHEP {\bf 1604}, 019 (2016)
  doi:10.1007/JHEP04(2016)019
  [arXiv:1512.00229 [hep-ph]].
  %%CITATION = doi:10.1007/JHEP04(2016)019;%%
  %2 citations counted in INSPIRE as of 08 Feb 2017
  
\end{thebibliography}
\end{document}